\documentclass[aps,prb,twocolumn,superscriptaddress,showpacs,floatfix,amsmath,amssymb]{revtex4-1}
\usepackage[breaklinks=true,colorlinks,citecolor=blue,linkcolor=blue,urlcolor=blue]{hyperref}
\usepackage{graphicx}
\usepackage{color}
\usepackage{ifthen}
\usepackage{amsmath}

\ifthenelse{\isundefined{\vec}}{ \newcommand{\vec}[1]{ \boldsymbol{ #1 } } }{ \renewcommand{\vec}[1]{ \boldsymbol{ #1 } } }
\newcommand{\mat}[1]{ \underline{\vec{#1}} }

\begin{document}

\title{Transient charge and energy flow in the wide-band limit}

\author{F. Covito}

\author{F. G. Eich}
\email[]{florian.eich@mpsd.mpg.de}

\author{R. Tuovinen}

\author{M. A. Sentef}
\affiliation{Max Planck Institute for the Structure and Dynamics of Matter, Luruper Chaussee 149, 22761 Hamburg, Germany}

\author{A. Rubio}
\affiliation{Max Planck Institute for the Structure and Dynamics of Matter, Luruper Chaussee 149, 22761 Hamburg, Germany}
\affiliation{Center for Free-Electron Laser  Science, Luruper Chaussee 149, 22761 Hamburg, Germany}
\affiliation{Center for Computational Quantum Physics (CCQ), The Flatiron Institute, 162 Fifth avenue, New York NY 10010}
\affiliation{Nano-Bio Spectroscopy Group, Departamento de Fisica de Materiales, Universidad del Pa\'is Vasco, 20018 San Sebasti\'an, Spain}

\begin{abstract} 
  The wide-band limit is a commonly used approximation to analyze transport through nanoscale devices. In this work we investigate its applicability to the study of charge and heat transport through molecular break junctions exposed to voltage biases and temperature gradients. We find that while this approximation faithfully describes the long-time charge and heat transport, it fails to characterize the short-time behavior of the junction. In particular, we find that the charge current flowing through the device shows a discontinuity when a temperature gradient is applied, while the energy flow is discontinuous when a voltage bias is switched on and even diverges when the junction is exposed to both a temperature gradient and a voltage bias. We provide an explanation for this pathological behavior and propose two possible solutions to this problem.
\end{abstract}

\maketitle

\section{Introduction} \label{SEC:Intro}
Over the last decades great effort has been spent to miniaturize electric circuits. The goal is to realize the fundamental building blocks of electronic circuits, such as transistors, on the scale of single molecules. There has been a great success in shrinking electronic devices down experimentally. In order to understand the properties of molecular break junctions a quantum mechanical description of the device is required. Perhaps the most successful and wide-spread theory to describe how charge flows through a nanoscale junction is the so-called Landauer-B\"uttiker approach \cite{Landauer:57,BuettikerPinhas:85,Landauer:89}, which describes the charge transport as a scattering problem. Essentially, the flow of charge through a molecular junction is determined by the transmission function of the device--describing how impinging electrons are scattered--and the occupation function of the electrons in the (metallic) leads connected to the junction. 

In recent years there has been renewed interest in addressing not only the charge flow, but also the energy (or heat) flow through nanoscale devices. Understanding how charge and energy flow depend on voltage and temperature biases across the device provides crucial insight for the development of thermoelectric circuits, which could be used to convert waste heat into useful electric energy~\cite{Goldsmid:09,DubiDiVentra:09}. Furthermore, recent experiments demonstrate that local temperatures in nanoscale conductors can be measured with a spatial resolution of tens of nanometers~\cite{MecklenburgRegan:15,HalbertalZeldov:16}. A common path to address the effect of temperature gradients across the nanoscale device is to allow for different temperatures in the occupation functions characterizing the leads in the Landauer-B\"uttiker formula. Conceptually this can only be justified if the leads are considered to be disconnected from the device initially (partitioned approach). This artificial partitioning of the system, however, is problematic, for it assumes that it is possible to perfectly decouple the leads from the molecular junction--a rather optimistic assumption if one considers atomic-scale devices. For times much larger than the typical time-scale of molecular break junctions, which are on the order of femtoseconds~\cite{KinoshitaMunakata:95,GauyacqRaseev:01,KirchmannWolf:05,ChulkovEchenique:06}, the assumption on a decoupled initial state does not play a crucial role. However, for transient dynamics the initial state matters. As pump-probe experiments are now able to investigate phenomena happening at this timescale~\cite{MyllyperkioPettersson:10,CockerHegmann:13,NiBasov:16,KarnetzkyHolleitner:17arxiv} it is important to properly describe the initial state.

An alternative to the partitioned approach is to couple the device and leads at all times and trigger the charge flow by switching a potential bias~\cite{Cini:80}. This \emph{partition-free} approach leads to the same steady state as the partitioned approach, but the transient dynamics of the device will, in general, be different~\cite{StefanucciAlmbladh:04}. The advantage of the partition-free approach is that the transient charge and energy/heat flows are not spoiled by the dynamics induced by connecting leads and device, because the electronic states in the device are allowed to hybridize with the leads before any temperature or voltage bias is applied. Importantly, it is also possible to take into account temperature differences in the leads within the partition--free approach: We consider a \emph{thermo-mechanical potential}, which couples to the \emph{local} energy density of the system--much like the usual electric potential couples to the charge density~\cite{Luttinger:64a}. This thermo-mechanical potential acts as mechanical ``proxy''~\cite{Shastry:09} for local temperature variations. An intuitive way to understand this is to consider the occupation function, which is determined from the ratio of the energy and the temperature. Accordingly, a change in occupations due to a change in temperature can alternatively be viewed as a change in energy keeping the temperature fixed. The thermo-mechanical potential allows to rescale the energy locally, thereby mimicking a locally varying temperature. Applying this idea in the context of transport means that different temperatures in the leads are described by rescaling the bandwidth of the leads~\cite{EichVignale:14b}.

A wide-spread simplification used to describe transport through nano junctions is the so-called wide-band limit (WBL). The WBL assumes that the detailed structure of the density of states in the leads is not important for the description of transport, which substantially simplifies computations. The WBL for charge transport is justified when the bandwidth is large compared to the applied bias~\cite{ZhuWang:05,MaciejkoGuo:06,ZhengChen:07,ZhangChen:13,VerzijlThijssen:13,ShiJin:16,Baldea:16}.

In this work we investigate whether the WBL can be employed in conjunction with the thermo-mechanical potential. An immediate question that comes to mind is: What is the meaning of rescaling an infinite band? In the following we will show that the steady state is well described in the WBL, provided the WBL is taken properly. The transient currents, however, exhibit peculiarities at short times. Specifically, we see that the charge current jumps at the initial time when the device is exposed to a temperature gradient and, similarly, the heat current behaves discontinuously when a voltage bias, but no temperature bias, is switched on. Even more dramatically, the heat current diverges as $(t-t_0)^{-1}$, with $t_0$ being the time at which a temperature \emph{and} charge bias is applied to the system. By comparing the WBL transient charge and heat currents to results obtained at finite bandwidth, we highlight that this pathological behavior of the WBL can be attributed to the fact that--at short times--the natural cut-off, provided by the finite bandwidth, plays a crucial role for the dynamics.
 
\section{Model and method} \label{SEC:Hamiltonian}
\begin{figure}
  \includegraphics[width=\columnwidth]{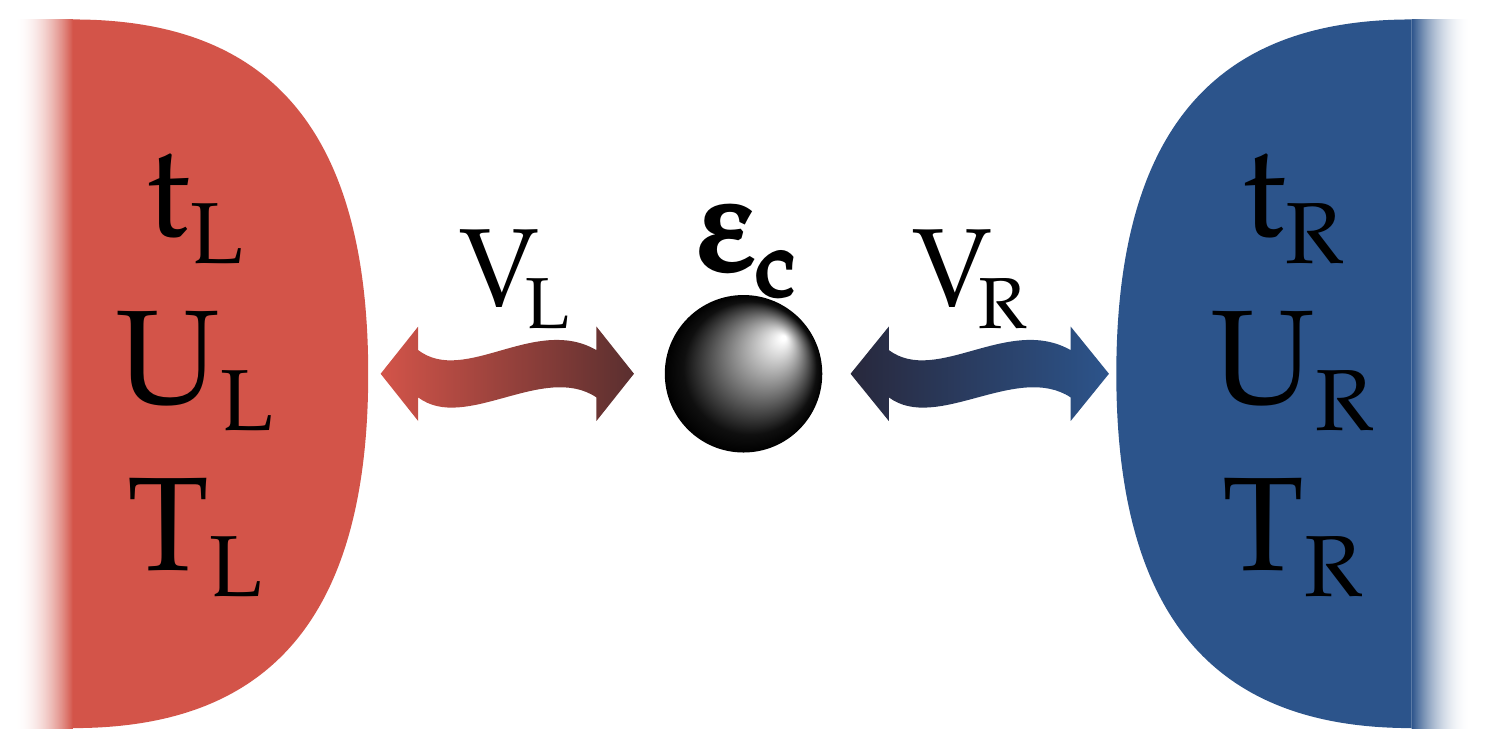}
  \caption{Molecular break junction: Schematic representation of the system considered in this work. A single impurity site, representing a molecular level with energy $\epsilon_c$, is coupled via hopping amplitudes $V_\alpha$ to metallic leads with a bandwidth $4t_\alpha$ ($\alpha=\mathrm{R},\mathrm{L}$). Charge and energy flow is triggered by applying a potential bias $U_\alpha$ to the leads and/or changing the temperature $T_\alpha$ in the leads. \label{FIG:sys}}
\end{figure}
We consider a simple tight-binding model Hamiltonian to describe a molecular break junction. A single molecular level is connected to two metallic leads (cf.\ sketch in Fig.\ \ref{FIG:sys}). The Hamiltonian reads
\begin{align}
  \hat{H} & = \epsilon_c \hat{\phi}^\dagger_{c} \hat{\phi}_{c} + 
  \sum_{\alpha k} \epsilon_{\alpha k} \hat{\phi}^\dag_{\alpha k} \hat{\phi}_{\alpha k} \nonumber \\
  & \phantom{=} {} + \sum_{\alpha k} \left( \hat{\phi}^\dag_{\alpha k} V_{(\alpha k) c} \hat{\phi}_{c} + \hat{\phi}^\dag_{c} V_{c (\alpha k)} \hat{\phi}_{\alpha k} \right) ~, \label{H}
\end{align}
where $\epsilon_c$ is the energy of molecular level, $\hat{\phi}^\dag_{\alpha k}$ and $\hat{\phi}_{\alpha k}$ are the field operators of the leads, with $\alpha=\text{L(eft)},\text{R(ight)}$ and $k$ labels the basis functions in the leads, and $\hat{\phi}^\dag_{c}$, $\hat{\phi}_{c}$ represent the field operators associated to the molecular level. The matrix elements $V_{(\alpha k) c} = [V_{c (\alpha k)}]^\star$ take the coupling between the molecular level and the leads into account. The leads are modeled as non-interacting one-dimensional tight-binding chains, i.e., the dispersion of the electrons in the leads is given by
\begin{align}
  \epsilon_{\alpha k}=-2t_\alpha \cos(k) + c_{\alpha}~, \label{eak}
\end{align}
where $t_\alpha$ is the nearest neighbor hopping in lead $\alpha$, yielding a bandwidth of $4t_\alpha$. The energy $c_\alpha$ corresponds to the center of the band of the lead $\alpha$, i.e., it determines the alignment of the band with respect to the chemical potential, which we take to be at zero energy. Finally the hopping to the central site is $V_{\alpha k} = V_\alpha \sin(k)$. The embedding self-energy due to lead $\alpha$ is then given by
\begin{align}
  \Sigma^{\mathrm{R}/\mathrm{A}}_\alpha(z) & =  \sum_k V_{(\alpha k) c} g^{\mathrm{R}/\mathrm{A}}_{\alpha k}(z)
  V^\star_{(\alpha k) c} \nonumber \\ 
  & = \frac{|V_\alpha|^2}{t_\alpha} S\left(\frac{z-c_\alpha}{2 t_\alpha}\right) ~, \label{Sigma}
\end{align}
with $g^{\mathrm{R}/\mathrm{A}}_{\alpha k}(z)$ being the retarded/advanced Green's function of the isolated lead $\alpha$. The function $S(z)$ is given by
\begin{align}
  S(z) = z - \sqrt{z-1}\sqrt{z+1} ~, \label{S}
\end{align}
where the character of the function $S(z)$, i.e., whether it is the advanced or retarded self-energy, is determined by the sign of the imaginary part of $z$~\footnote{The function $S(z)$ has a branch cut on the real axis from $z=-1 \to z=1$.}. The Green's function for the molecular level is then simply given by
\begin{align}
  G^{\mathrm{R}/\mathrm{A}}(z) = \left[z - \epsilon_c
  - \sum_{\alpha} \Sigma^{\mathrm{R}/\mathrm{A}}_\alpha(z) \right]^{-1} ~. \label{Gdevice}
\end{align}
The inverse of the imaginary part of the self-energy yields a finite lifetime for the quasi-particles in the molecular junction, and the real part of the self-energy shifts the energy of the quasi-particles. 

The WBL is defined as the limit $t_\alpha\rightarrow\infty$ (infinite bandwidth) while keeping the ratio $|V_\alpha|^2/t_\alpha$, which corresponds to the decay rate into lead $\alpha$, constant. Expanding the expression of the self-energy for large $t_\alpha$ we obtain
\begin{align}
  \left[\Sigma^{\mathrm{R/A}}_\alpha(z)\right]_{\mathrm{WBL}} = \mp i \frac{|V_\alpha|^2}{t_\alpha} = \mp i \frac{\Gamma_\alpha}{2} ~, \label{SigmaWBL}
\end{align}
where the $\mp$ sign refers to the retarded/advanced self-energy, respectively. As we can see from this expression, the only effect of the leads is to provide a decay-mechanism for the quasi-particles.

Expressing the field operators $\hat{\phi}$ in the Heisenberg picture and using their equations of motion, the charge and heat currents are given by~\cite{EichVignale:14b,ArracheaMartinMoreno:07,EspositoGalperin:15b,LudovicoSanchez:16}:
\begin{subequations} \label{IQ}
  \begin{align}
    I_\alpha & = - \partial_t \sum_k \langle \hat{\phi}^\dag_{\alpha k}(t)
    \hat{\phi}_{\alpha k}(t)\rangle ~, \label{Ia} \\
    Q_\alpha & = -\partial_t \left[\sum_k \epsilon_{\alpha k}
    \langle \hat{\phi}^\dag_{\alpha k}(t) \hat{\phi}_{\alpha k}(t)\rangle \right.
    ~\label{Qa} \\
    & \phantom{=} \; \left. {} + \frac{1}{2} \sum_{k n} \bigg( V_{(\alpha k) n}
    \langle \hat{\phi}^\dag_{\alpha k}(t) \hat{\phi}_{n}(t)\rangle + \text{h.c.}\bigg) \right].
    \nonumber
  \end{align}
\end{subequations}
Note that we define the heat current $Q_\alpha$ as the temporal change in the energy within the leads \emph{plus} half the coupling energy\cite{LudovicoSanchez:14}.

\section{Transport setup} \label{SEC:Bias}
\begin{figure}
  \includegraphics[clip,width=.45\textwidth]{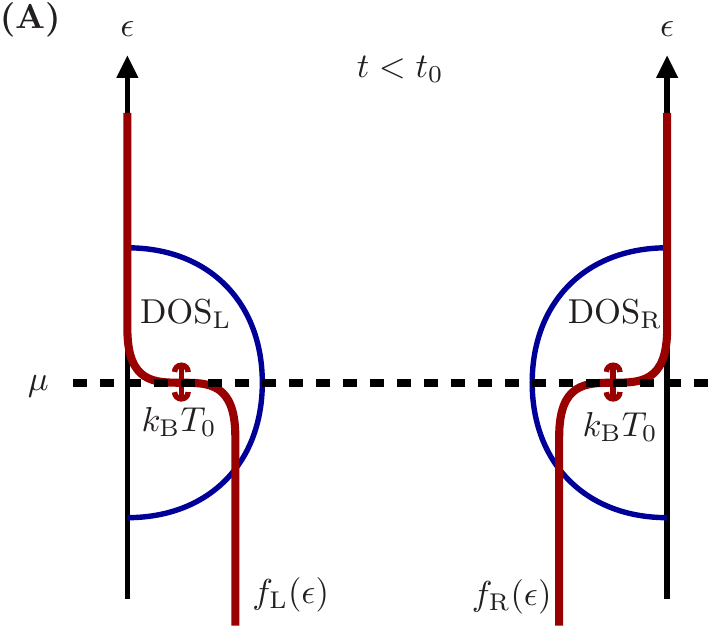}
  \includegraphics[clip,width=.45\textwidth]{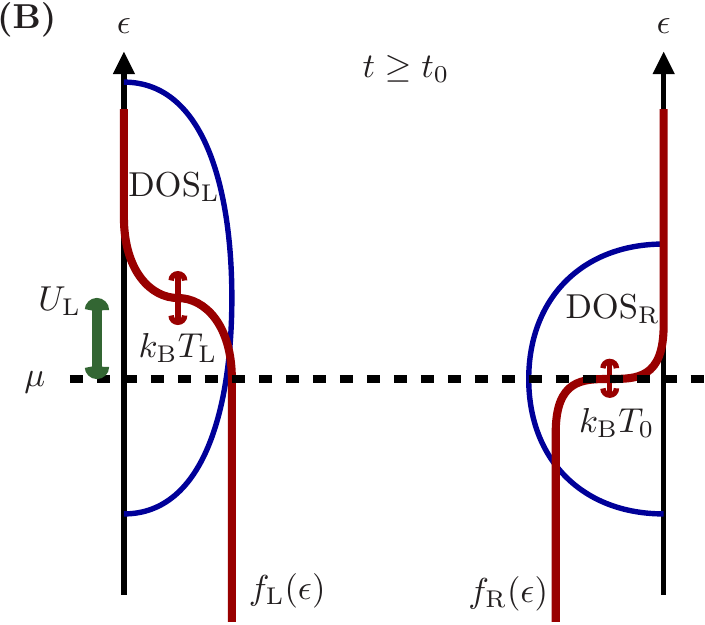}
  \caption{Comparison of initial and steady state: (A) Initial state of the junction. The leads--molecule system is equilibrated at a unique temperature $T_0$ and chemical potential $\mu$ (represented by the dashed horizontal line). (B) Graphical representation of the steady state. At $t_0$ a thermo-mechanical potential and the potential bias is applied. This results in a steady state in which the occupation function of the left lead corresponds to a Fermi function with $T_\mathrm{L} = 2 T_0$ and $\mu_\mathrm{L} = \mu + U_\mathrm{L}$.\label{FIG:TMsketch}}
\end{figure}
In this work, we will investigate the validity of the WBL in the case of dynamical heat and charge transport in the junction described in the previous section. Once a non-equilibrium situation is created by applying a potential bias and/or temperature gradient, transient dynamics will take place and electrons will move, resulting in charge and heat currents flowing across the junction. We focus on the specific cases of quenches, i.e., the electric and thermo-mechanical potentials suddenly change at a certain time $t_0$. Transient dynamics, induced by changing the potentials, occur on the order of a characteristic time scale $\tau$ given by the inverse of the decay rate provided due to the leads, i.e., 
\begin{align}
  \tau^{-1} = \sum_{\alpha} \frac{V_\alpha^2}{t_\alpha} ~. \label{tau}
\end{align}
For times $t \gg \tau$ the junction will reach a steady state. We choose the hopping $V_\mathrm{L} = V_\mathrm{R} = V$ as our unit of energy: the molecular energy level is taken to be at $\epsilon_c = 0.2 V$, the nearest neighbor hopping of the leads $t_\mathrm{L}=t_\mathrm{R}=5V$, the chemical potential defines the zero of the energy, the center of the bands of the leads are aligned with it $(c_\alpha=\mu=0)$, and the (inverse) temperature $\beta = (k_\mathrm{B} T_0)^{-1} = 100 V^{-1}$. 

\begin{figure}
  \centering
  \includegraphics[width=\columnwidth]{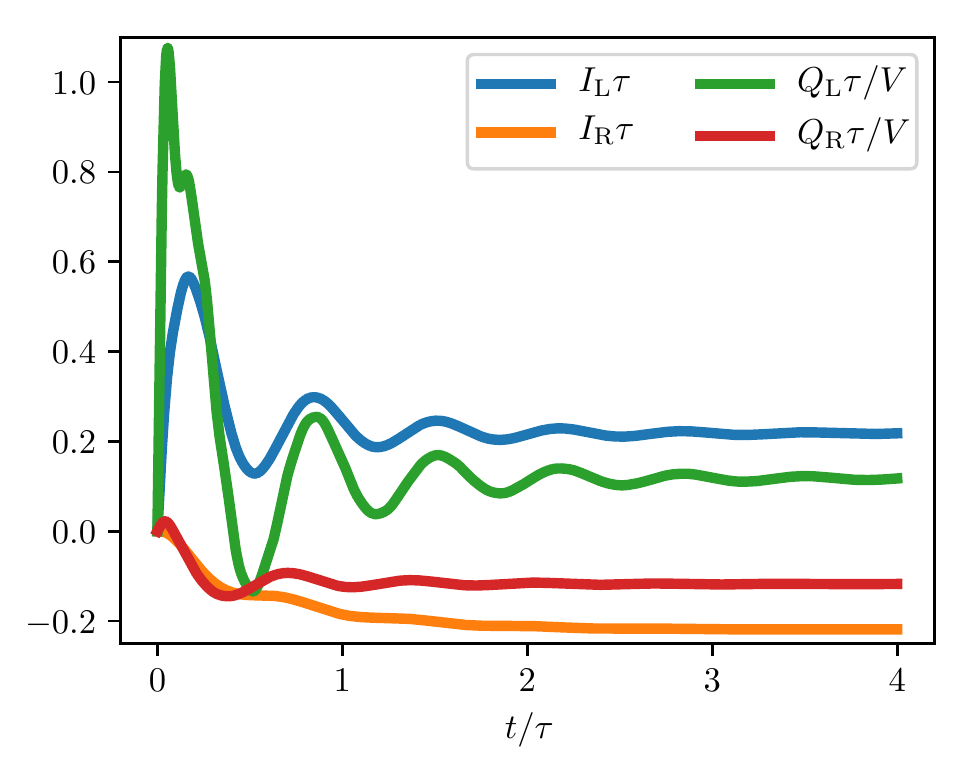}
  \caption{Charge and heat currents: Transient charge ($I_\alpha$) and heat currents ($Q_\alpha$) flowing between the left lead ($\alpha=\mathrm{L}$) and the right lead ($\alpha=\mathrm{R}$) and the molecular junction, respectively. The results are obtained taking into account the full frequency dependence of the embedding self-energy. The currents are triggered by a sudden change in the temperature and potential in the left lead at $t_0=0$.}
  \label{FIG:IQtransients}
\end{figure}

For $t<t_0$ the system is taken to be in thermal equilibrium at temperature $T_0$. In order to induce a charge current through the junction the left lead is shifted up in energy by $U=2V$ for $t\geq t_0$, i.e., the energy dispersion of the left lead is given by
\begin{equation}\label{potentialBias}
\epsilon_{\mathrm{L} k} = \begin{cases}
-2t_\mathrm{L}\cos(k) & \text{for} \ t<t_0, \\ -2t_\mathrm{L}\cos(k) + U & \text{for} \ t \geq t_0.
\end{cases}
\end{equation}
In order to describe a temperature gradient across the junction--in addition to the potential bias--we apply a thermo-mechanical potential $\psi = \frac{T_\alpha - T_0}{T_0} = 1$ in the left lead, which rescales the bandwidth for $t\geq t_0$. This thermo-mechanical potential effectively doubles the temperature in the left lead:
\begin{equation}\label{chargeTemperatureBias}
\epsilon_{\mathrm{L} k} = \begin{cases}
-2t_\mathrm{L}\cos(k), & \text{for} \ t<t_0 \\ (1+\psi)(-2t_\mathrm{L}\cos(k) + U) & \text{for} \ t \geq t_0.
\end{cases}
\end{equation}
Figure \ref{FIG:TMsketch} sketches of the molecular junction in the initial equilibrium and in the steady-state limit, showing that in the steady-state the energy dispersion of the left lead is broadened by a factor of two. In Fig.\ \ref{FIG:IQtransients} we depict the time-dependent charge and heat currents through the molecular junction. In this calculation, both, a charge bias and a temperature gradient is applied across the junction and we observe fast transient oscillation of the currents on the timescale $\tau$ followed by a saturation to a steady current.

\section{Results} \label{SEC:Results}
In order to test the WBL we compute the time-dependent charge and heat currents flowing from the leads into the impurity in the WBL and compare the results to calculations taking the full frequency dependence of the lead self-energy [cf.\ Eq.\ \eqref{S} and Fig.~\ref{FIG:IQtransients}] into account. Specifically, we rescale the bandwidth of the leads, making it effectively wider, while keeping the ratio $|V_\alpha|^2/t_\alpha$ constant. Hence, we use
\begin{align}
  V^\lambda_\alpha = \sqrt{\lambda}V_\alpha \;\;\;,\;\;\;
  t^\lambda_\alpha =\lambda t_\alpha ~, \label{rescaling}
\end{align}
with a rescaling factor $\lambda$, which allows us to approach the WBL as $\lambda \to \infty$. 
We focus on two different scenarios: 1) A situation where only a potential bias is applied to the left lead [cf.\ Eq.\ \eqref{potentialBias}], 2) A situation where, both, a potential bias and a temperature difference are applied across the junction [cf.\ Eq.\ \eqref{chargeTemperatureBias}]. The numerical algorithm to compute the transient currents--taking the full frequency dependence of the embedding self-energy into account--has been already discussed in Ref.\ \onlinecite{EichVignale:16}. In the following we refer to these results as the ``full'' calculation. Very recently progress has been made in evaluating the time-dependent currents in tight-binding models within the WBL analytically~\cite{TuovinenVanLeeuwen:14,RidleyKantorovich:15}. It turns out that this is also possible if a thermo-mechanical potential--describing temperature gradients--is present. Accordingly, all WBL results are obtained analytically. The explicit derivation of the analytical expression will be presented elsewhere.

\emph{Steady state currents.} For times much longer than the characteristic lifetime $\tau$ the system reaches a steady state. In general we find that the steady state currents obtained in the WBL coincides with the results of the full calculation when the scaling factor $\lambda$ is increased. However, there is a subtle point in the evaluation of the heat current in the steady state: it turns out to be crucial to take the WBL at the end of the calculation and \emph{not} inside the integral defining the steady state current. The difference between taking the WBL inside the integral and taking the WBL after performing the integral is only present if two leads at different temperatures are connected to the same state in the device. This is trivially the case for a molecular junction modeled by a single site. We present a careful derivation in App.\ \ref{APP:WBL} showing that the order of limits matters.

\begin{figure}
  \includegraphics[width=\columnwidth]{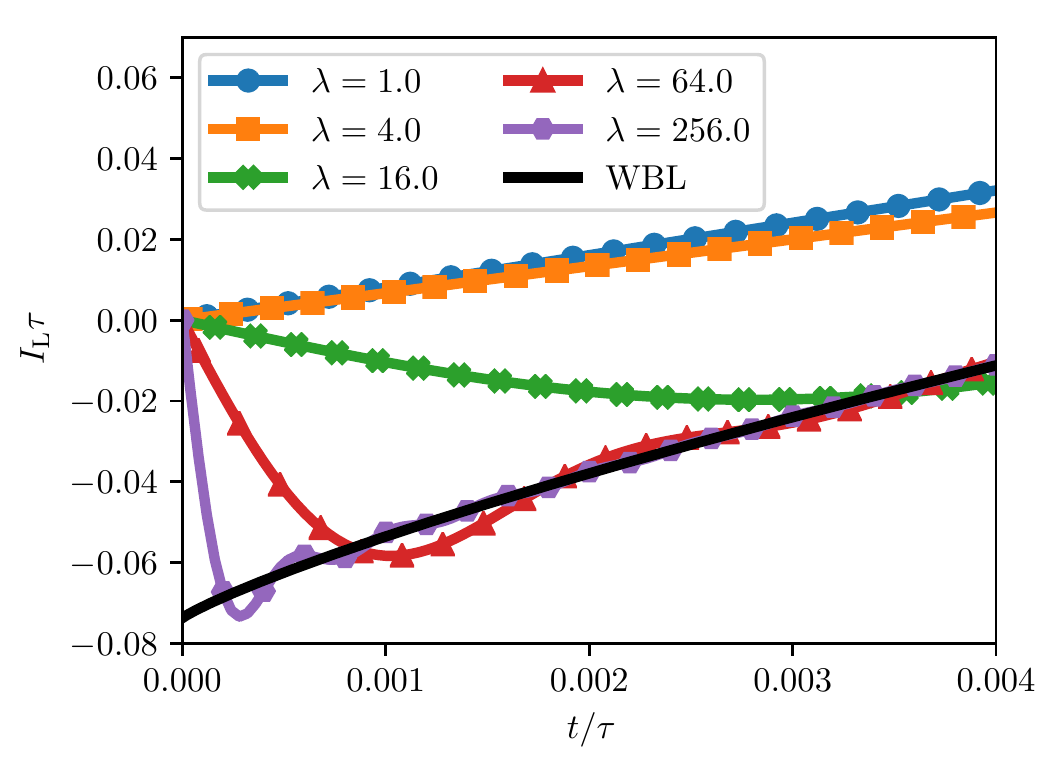}
  \caption{Short-time dynamics of charge current: Time-dependent charge current flowing from the left lead to the impurity under the influence of a potential bias and a temperature gradient. The full calculation approaches the WBL result as the scaling factor increases. However, there is an apparent discontinuity developing at $t=0$ for $\lambda \to \infty$. While in the full calculation the charge current always vanishes for $t \to 0$, in the WBL a finite value is obtained. \label{FIG:IL_dT}}
\end{figure}
\emph{Transient charge current.} The transient charge induced by a potential bias alone is nicely reproduced in the WBL [cf.\ App.\ \ref{APP:TransientPlots} for the corresponding plots]. If a temperature gradient--in addition to the potential bias--is applied the charge current exhibits a jump at the initial time, but otherwise represents the full calculation for times $t \gtrsim \tau$. In Fig.\ \ref{FIG:IL_dT} we depict the charge current for $t \ll \tau$ (a plot of $I_\alpha$ for $t \gtrsim \tau$ is provided in App.\ \ref{APP:TransientPlots}). It can be shown analytically that the jump, $\Delta I_\alpha$, at the initial time is proportional to
\begin{align}
  \Delta I_\alpha \propto \bar{V}_\alpha \left( \frac{V_\alpha}{t_\alpha}
  - \frac{\bar{V}_\alpha}{\bar{t}_\alpha} \right) ~, \label{DIa}
\end{align}
where the hopping amplitude inside the leads for $t<t_0$ is denoted by ${t}_\alpha$ and for $t\geq t_0$ by $\bar{t}_\alpha$. Similarly, we could write the coupling between lead $\alpha$ and the molecular region as ${V}_\alpha$ before $t_0$ and $\bar{V}_\alpha$ afterwards. In our setup the couplings are held constant at all times, i.e., $\bar{V}_\alpha = V_\alpha$, and the temperature gradient is mimicked by changing the hopping inside the leads as discussed in Sec.\ \ref{SEC:Bias}. Specifically, from Eq.\ \eqref{chargeTemperatureBias} we have $\bar{t}_\mathrm{L} = (1 + \psi) t_\mathrm{L}$, which means that the charge current has a finite jump when a temperature gradient is applied across the molecular junction. Equation \eqref{DIa}, however, suggests that the jump can be avoided if the temperature bias is mimicked by scaling the couplings $V_\alpha$ in the same way as the hopping inside the leads. This would imply that $\bar{V}_\alpha / \bar{t}_\alpha = V_\alpha / t_\alpha$, which is sufficient to make $\Delta I_\alpha$ vanish even in the presence of a temperature gradient. We stress that this cannot be achieved in the partitioned approach, because $V_\alpha$ is zero by definition for $t<t_0$ if the system is initially decoupled.

\begin{figure}
  \includegraphics[width=\columnwidth]{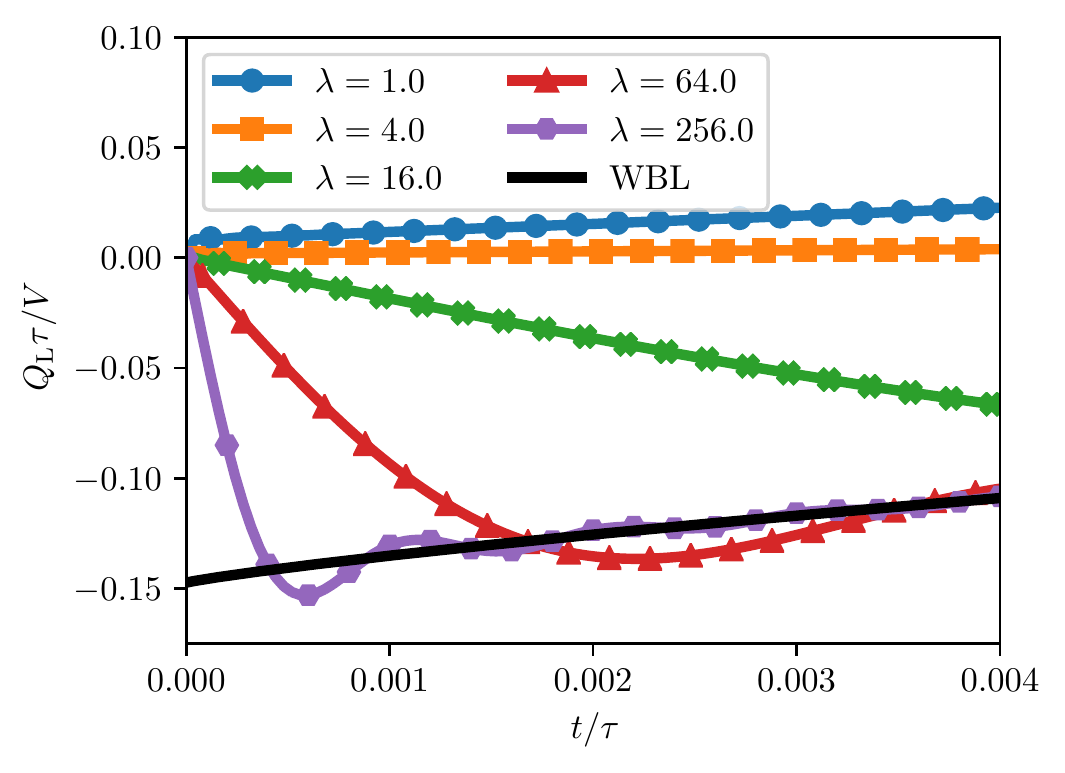}
  \caption{Transient heat current driven by a potential bias: Similar to the case of a charge current driven by a temperature gradient [cf.\ Fig.\ \ref{FIG:IL_dT}] in the limit of infinite bandwidth, $\lambda \to \infty$, the heat current develops a step at $t_0$. This means that the heat currents in the WBL tend to a finite value.}
  \label{FIG:QL_dU}
\end{figure}
\emph{Transient heat current.} Turning to the transient heat current we find that if only a potential bias is applied the heat current of the full calculation is reproduced in the WBL for times $t \gtrsim \tau$, but exhibits a jump at $t_0$. The short time behavior, $t \ll \tau$ is depicted in Fig.\ \ref{FIG:QL_dU} (cf.\ App.\ \ref{APP:TransientPlots} for $t\gtrsim \tau$). Similar to the case of the charge current induced by a temperature gradient, we can see that the WBL approximates a discontinuity at $t_0$ in the limit $\lambda \to \infty$ in the full calculation. Again, the heat current in the full calculation always vanishes as $t \to t_0$, but the WBL leads to a finite step in the heat current already in the presence of only a potential bias. In contrast to the charge current it is not possible to extract a simple expression as Eq.\ \eqref{DIa}, but instead the jump depends on the details of the molecular junction, i.e., on the quasi-particle energy levels.

\begin{figure}
  \includegraphics[width=\columnwidth]{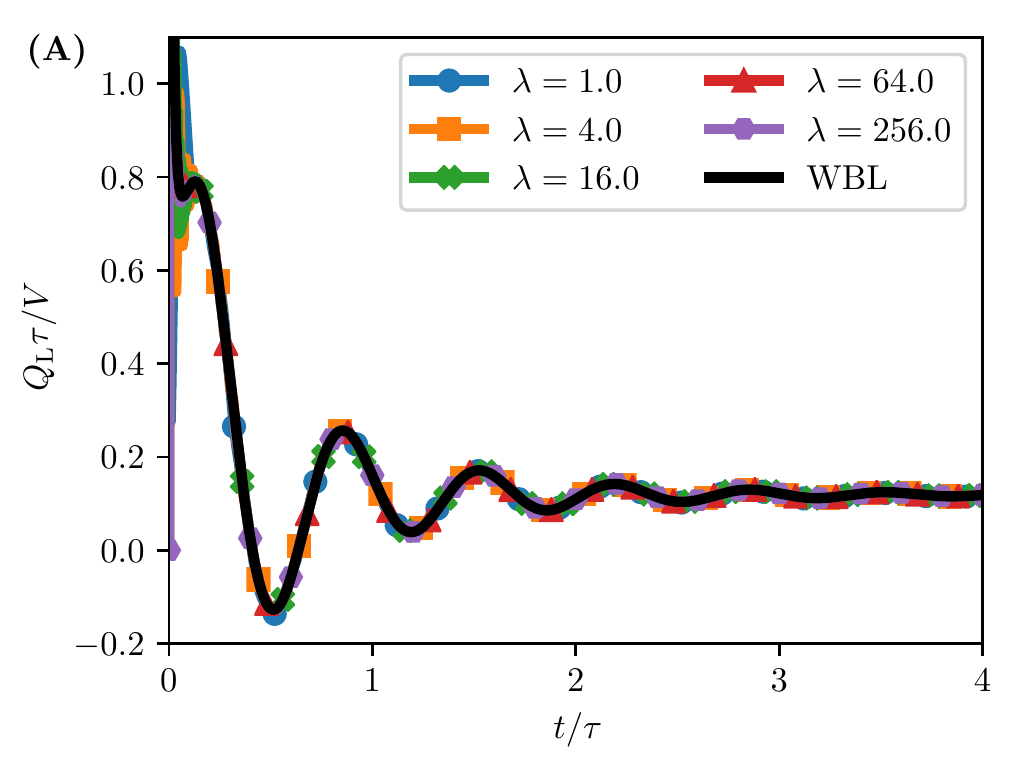}
  \includegraphics[width=\columnwidth]{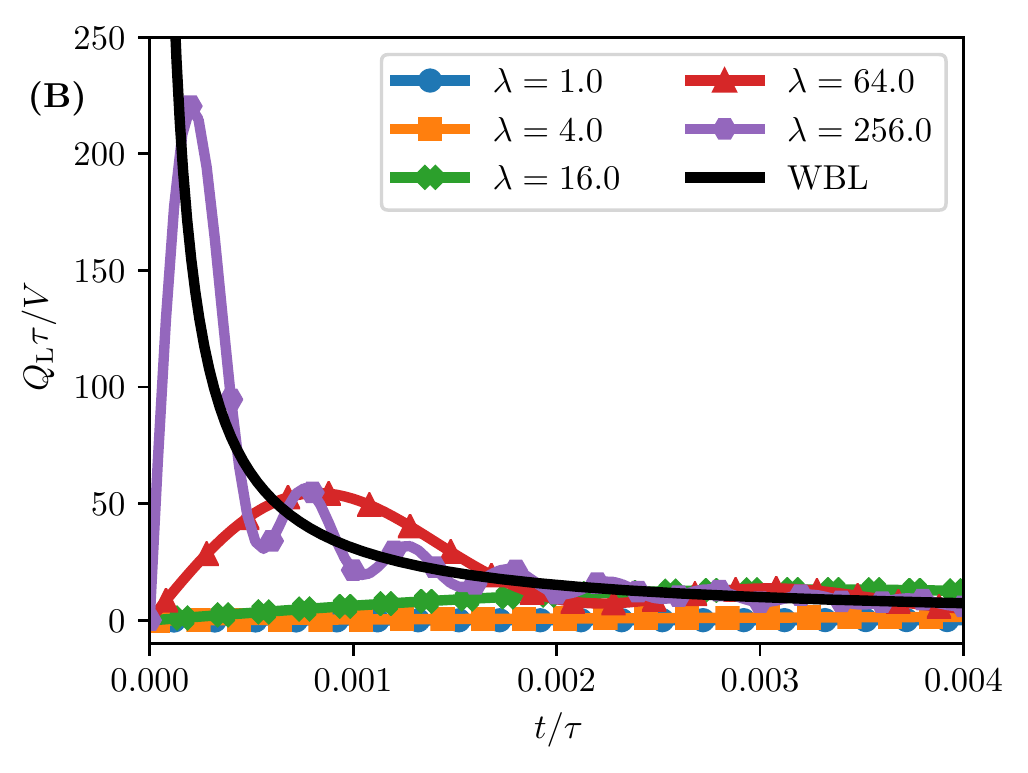}
  \caption{Heat current induced by temperature gradient: (A) Transient heat current for time comparable to the quasi-particle lifetime driven by a potential bias and a temperature gradient. For $t \gtrsim \tau$ the WBL reproduces the full calculations. (B) Transient heat current for $t \ll \tau$. The heat current in the full calculation exhibits a sharp spike which increases in height, and gets closer to $t_0$ as the bandwidth, which is proportional to $\lambda$ increases. The WBL results diverge as $(t-t_0)^{-1}$.}
  \label{FIG:QL_dT}
\end{figure}
If a temperature gradient is applied across the junction the heat current flowing from the left lead into the molecule depends strongly on the bandwidth for short times, even in the full calculation. In Fig.\ \ref{FIG:QL_dT} we can see that the heat current oscillates strongly for $t \lesssim \tau$ with a frequency proportional to the bandwidth (which, in turn, is proportional to $\lambda$). These oscillations correspond to transitions between the band edges of the leads and have been already observed in Ref.\ \onlinecite{EichVignale:16}. In the WBL these oscillations are absent since there are no band edges, but instead the heat current diverges as
\begin{align}
  Q_\alpha \sim \bar{V}_\alpha \left( \frac{{V}_\alpha}{{t}_\alpha}
  - \frac{\bar{V}_\alpha}{\bar{t}_\alpha} \right) \frac{1}{t-t_0} ~. \label{QaDivergence}
\end{align}
In the partition-free approach this divergence can be tamed by rescaling the couplings $V_\alpha$ in the same way as $t_\alpha$, i.e., by applying the thermo-mechanical potential not only inside the leads but also on the boundary of the junction. It turns out, however, that the sub-leading order for the heat current exhibits a logarithmic divergence as $t\to t_0$.
\begin{figure}
  \includegraphics[width=\columnwidth]{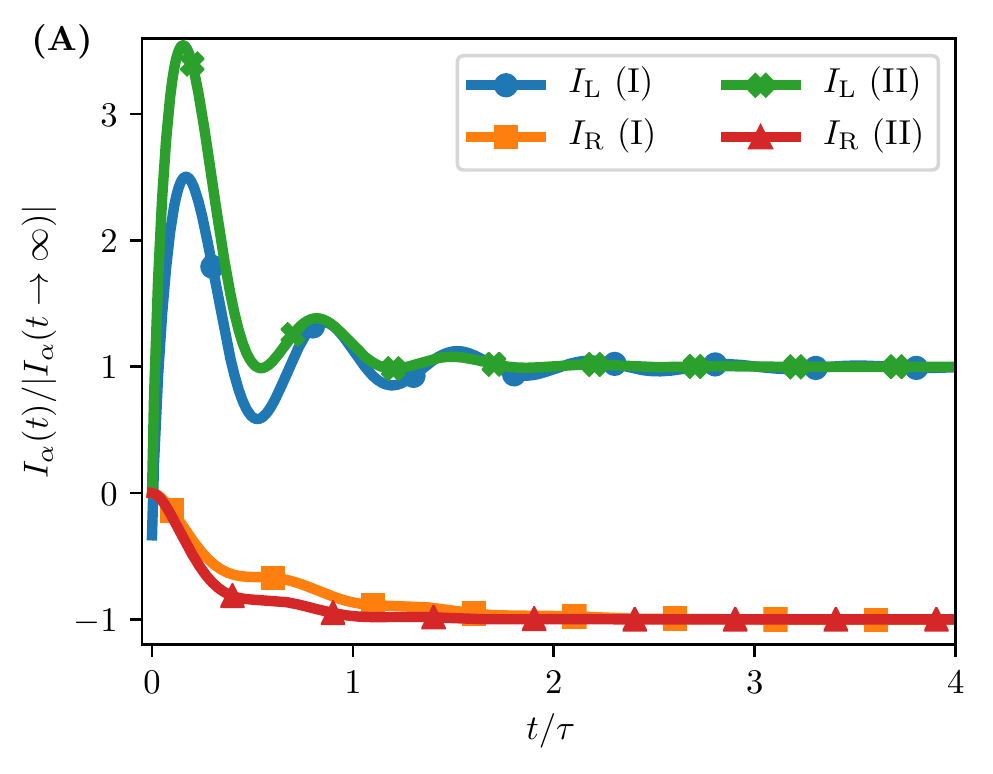}
  \includegraphics[width=\columnwidth]{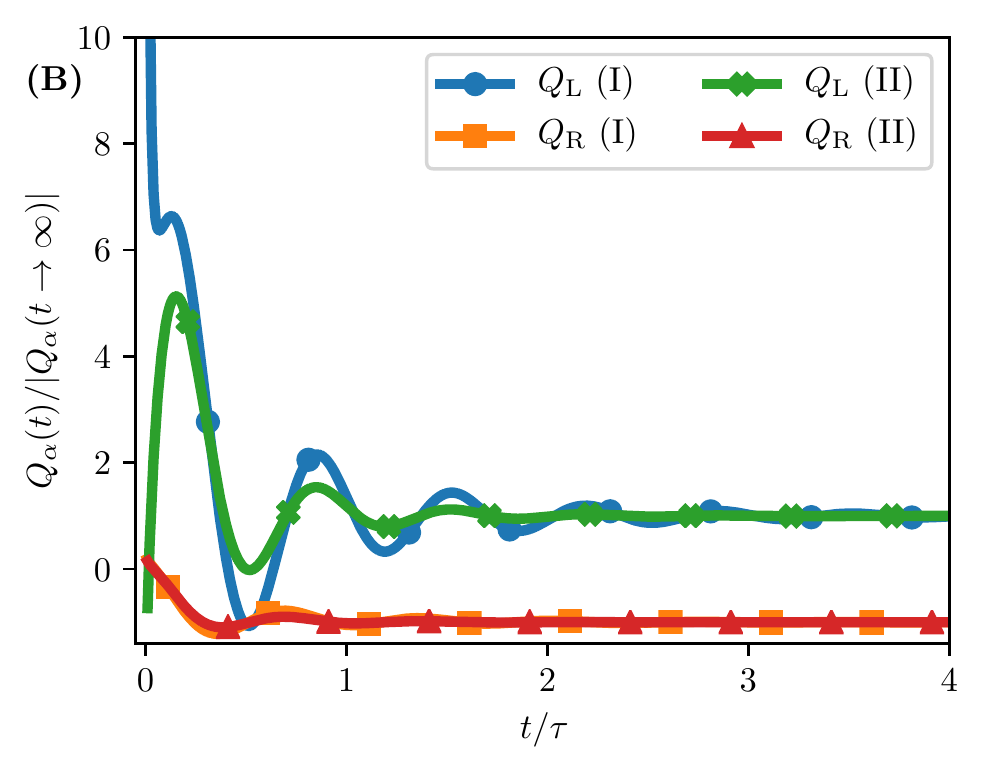}
  \caption{Fixing the short-time behavior of the currents: (A) Transient charge current for times comparable to the quasi-particle lifetime driven by a potential bias and a temperature gradient. (B) Transient heat current. In both panels we compare the WBL results obtained by applying the thermo-mechanical potential only inside the leads (I) to the WBL currents obtained by applying the thermo-mechanical potential also to the coupling between the leads and the device (II). The currents are normalized by their respective steady-state limit.}
  \label{FIG:IQ_dT_WBLcomp}
\end{figure}
This is shown in Fig.\ \ref{FIG:IQ_dT_WBLcomp}, where we compare the transient charge and heat currents in the WBL scaling only the hopping inside the leads (I) and scaling also the coupling to the impurity (II). We can see that the charge current starts from zero for both leads if also the coupling $V_\mathrm{L}$ is rescaled with the temperature, but it exhibits a finite jump in the lead where the temperature is changed if the temperature only rescales the hopping inside the lead. For the heat current we see that currents in all leads exhibit a logarithmic divergence as $t \to t_0$ if, both, $t_\alpha$ and $V_\alpha$ is rescaled due to the change in temperature. If only $t_\alpha$ is rescaled we see the aforementioned $(t - t_0)^{-1}$ divergence.

\section{Discussion and Conclusion} \label{SEC:Discussion}
In this work we have carefully examined the WBL for the transient and steady-state charge and heat currents through a molecular break junction. While we find that the long-time dynamics are faithfully captured in the WBL, at short times the WBL deviates considerably from a calculation taking the full frequency dependence of the embedding self-energy into account. This can be understood intuitively by considering that short times implies a wide spread in energy, and, therefore, the dynamics will be sensitive to whether the self-energies have a high frequency cut off (or decay) or whether they are constant for all frequencies. Specifically we have shown that the charge current induced by a temperature gradient, and the heat current induced by a potential bias, exhibit an unphysical jump at the initial time, when the system is suddenly quenched. Even more dramatically, the heat current diverges shortly after the quench if in addition to the potential bias also a temperature gradient is applied to the system. We have shown that these unphysical behaviors of the charge and heat current due to a temperature quench can be mitigated by considering that the temperature change not only affects the metallic leads, but also the boundary between the leads and the molecular junction. Since, in practice, this boundary is not sharply defined, we consider this a legitimate fix for the WBL. We stress that this fix can only be applied in the partition-free approach to the transport problem, i.e., when the coupling between the metallic lead and the molecular junction is already taken into account in the initial state of the system (before the quench). While this fix renders the charge current physical, in the sense that the initial current vanishes, the divergence in the heat current remains, but is only logarithmic. 

In order to address this, we see two possible solutions: 

1) In an actual experiment temperature gradients and potential biases will never be switched on infinitely fast, so a description as a sudden quench is questionable--to say the least--considering short time transient dynamics. It seems plausible that any kind of continuous switching will lead to a physical result (zero initial charge and heat currents). 

2) The second possible ``solution'' concerns the very definition of the energy or heat current between the leads and the device~\cite{ArracheaMartinMoreno:07,EspositoGalperin:15b,LudovicoSanchez:16}. In this work, the \emph{heat current} $Q_\alpha$ has been defined as the change in time of the \emph{internal} energy of the leads plus half of the coupling energy. Alternatively, the \emph{energy current} $J_\alpha$ from a certain lead $\alpha$ could be defined excluding the energy associated with the coupling to the device. The divergence at small times is due to the internal energy of the leads, which occurs in both the energy ($J_\alpha$) and the heat current ($Q_\alpha$). However there is yet another possible definition of the energy flowing between the leads and the device, i.e., we can define an energy current, $E_\alpha$, via the change of the energy stored in the device itself excluding the coupling. This leads to an expression $\partial_t \langle H \rangle = \sum_{\alpha} E_\alpha$, where $\langle H \rangle$ is the expectation value of the energy inside the junction. For a simple one-site model we have trivially $E_\alpha = \epsilon_c I_\alpha$, i.e., the new energy current is proportional to the charge current. However, for a multi-state device region, this is not necessarily the case.

We point out that from a numerical point of view it would be highly desirable to employ the WBL to compute transient charge and heat currents, because it affords an analytical solution in terms of the quasi-particle states and energies in the molecular device. Harnessing this analytical solution would allow for an efficient simulation of mesoscopic devices. An approach somewhat intermediate between taking the full frequency dependence of the self-energy into account and the WBL, could be to approximate the self-energies by Lorentzians, which provide a self-energy with the proper decay at high frequencies while allowing also for a (semi-)analytic solution of the transport problem~\cite{MaciejkoGuo:06,ZhangChen:13}.

In this work we only consider the non-interacting case. Through the Keldysh formalism~\cite{Keldysh:64_original,Keldysh:65,StefanucciVanLeeuwen:13} a generalization of the Landauer-B{\"u}ttiker formula for interacting electrons is also possible~\cite{MeirWingreen:92}. There is a very interesting alternative approach for tackling the interacting transport problem using time-dependent density-functional theory (TD-DFT)~\cite{RungeGross:84,Ullrich:12}, where the interacting problem is mapped onto a fictitious noninteracting problem. This implies that the Landauer-B\"uttiker formula applies. The effect of the electron-electron interaction is taken into account via an effective potential, which renormalizes the effective bias driving the charge flow~\cite{KurthGross:05,StefanucciRubio:07,KurthStefanucci:13}. Furthermore, the coupling to a thermo-mechanical potential can be used to generalize TD-DFT to allow for a direct description of \emph{charge and energy} flow~\cite{EichVignale:14a,EichVignale:17a}. We are confident that the combination of these approaches holds promise for studying the transient charge and energy flow in large molecular junctions for interacting systems.

\begin{acknowledgments}
F. G. E. has received funding from the European Union's Framework Programme for Research and Innovation Horizon 2020 (2014-2020) under the Marie Sk{\l}odowska-Curie Grant Agreement No. 701796. R. T. and M. A. S. acknowledge funding by the DFG through the Emmy Noether programme (SE 2558/2-1). A. R. acknowledges financial support from the European Research Council(ERC-2015-AdG-694097) and Grupos Consolidados (IT578-13).
\end{acknowledgments}

\begin{appendix}

\section{Additional plots for the transient currents}\label{APP:TransientPlots}

\begin{figure}
  \includegraphics[width=\columnwidth]{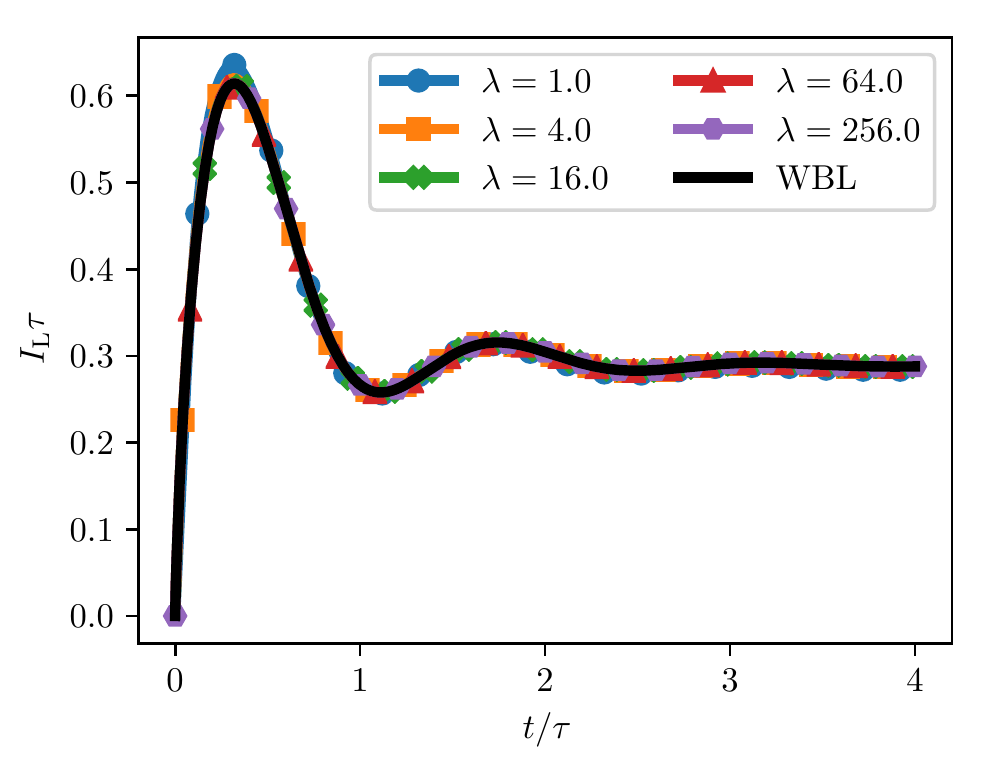}
  \includegraphics[width=\columnwidth]{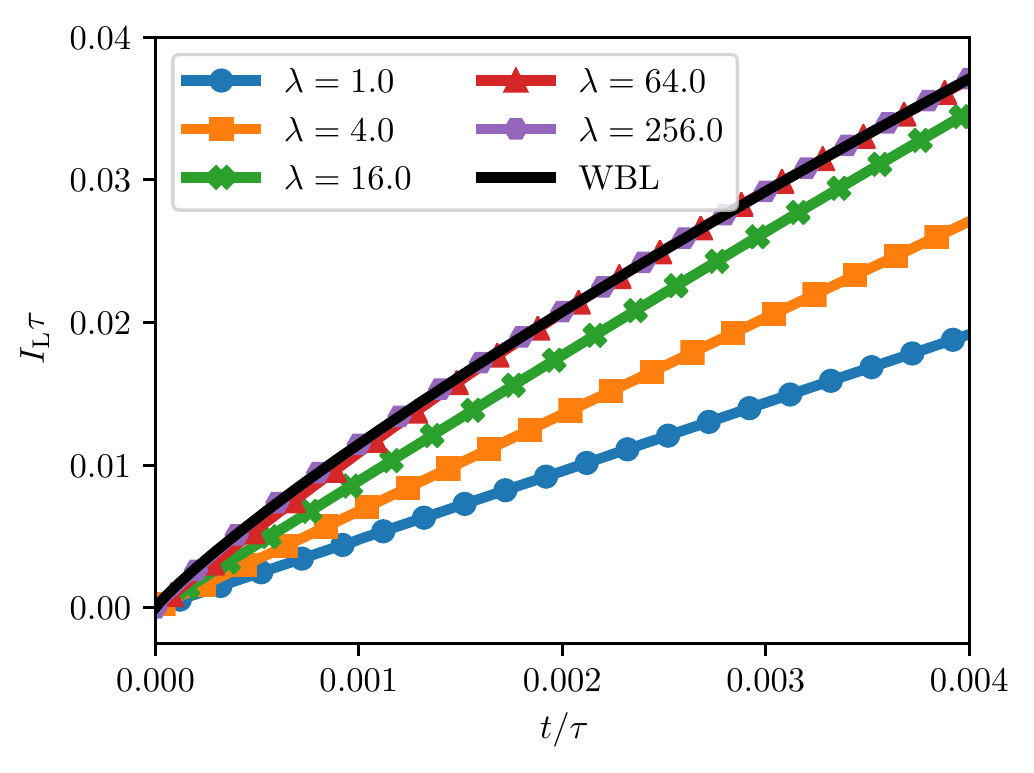}  
  \caption{Time-dependent charge current flowing from the left lead to the impurity when only a potential bias is applied. Upper panel shows the current for times comparable to $\tau$, while the lower panel depicts the transient current at $t \ll \tau$.}
  \label{FIG:I_dU}
\end{figure}

In this appendix we provide additional plots comparing the full results to the WBL. Figure \ref{FIG:I_dU} shows the transient charge currents for times comparable to the quasi-particle lifetime $\tau$ (upper panel) and for very short times (lower panel). It clearly shows that the WBL represents the $\lambda \to \infty$ limit of the full calculation.

\begin{figure}
  \includegraphics[width=\columnwidth]{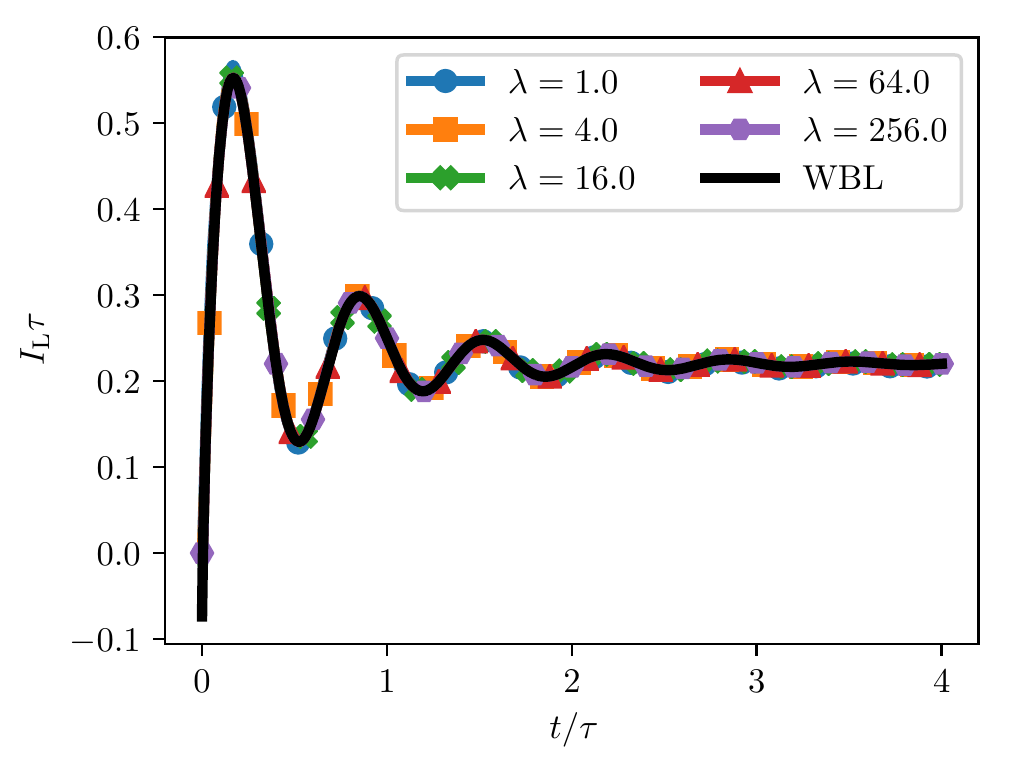}
  \includegraphics[width=\columnwidth]{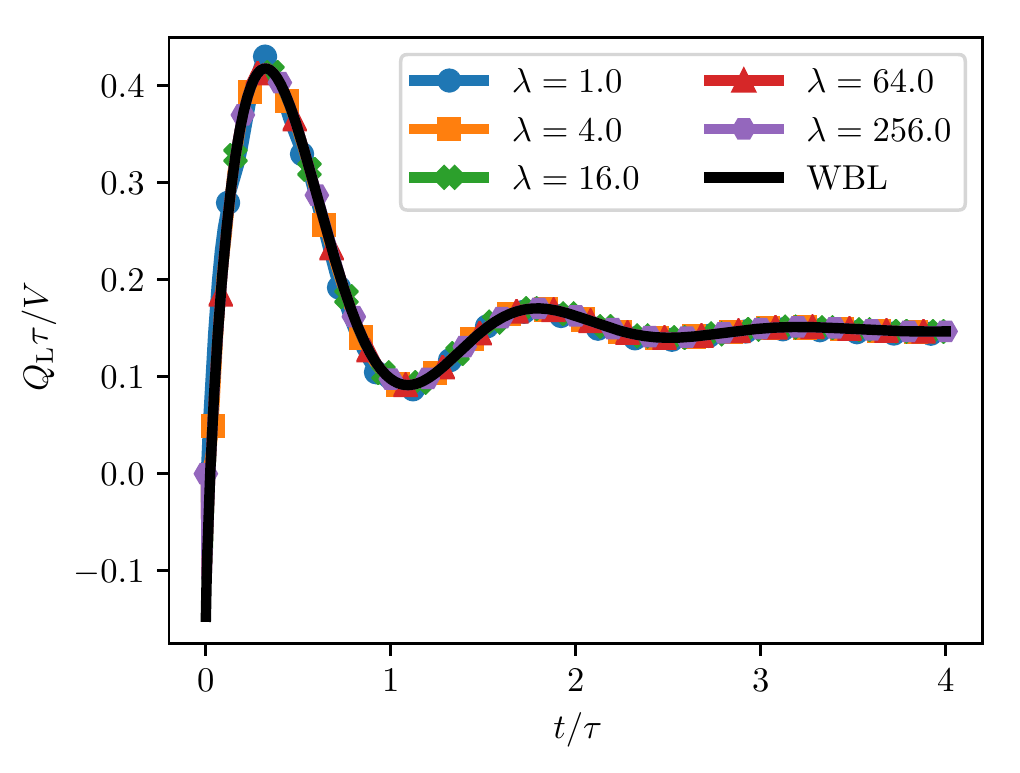}
  \caption{Upper panel shows the time-dependent charge current flowing from the left lead to the impurity when a potential bias and temperature gradient is applied across the junction. Lower panel shows the heat current induced by applying a potential bias only.}
  \label{FIG:IQ_LT}
\end{figure}

Fig.~\ref{FIG:IQ_LT} shows the charge and heat current for times comparable to the quasi-particle lifetime. The charge current is shown for the second scenario, i.e., when a potential bias and a temperature gradient is applied. The heat current is shown for the first scenario, i.e., when only a potential bias is applied at $t_0$. We see that in both cases the WBL currents correspond to the $\lambda \to \infty$ currents obtained in the full calculation for times $t\geq t_0$. However, in the WBL both currents approach a finite value for $t \to 0$ as discussed in Sec.\ \ref{SEC:Results}.


\section{Derivation of the steady-state energy current} \label{APP:WBL}
Here we present the analytical evaluation for the steady-state energy current discussed in Sec.~\ref{SEC:Results}. In our derivation we consider a generic Hamiltonian of the form
\begin{align}
  \hat{H} & = \hat{\vec{\Phi}}^\dagger \cdot \mat{H} \cdot \hat{\vec{\Phi}} + 
  \sum_{\alpha k} \epsilon_{\alpha k} \hat{\phi}^\dag_{\alpha k} \hat{\phi}_{\alpha k} \nonumber \\
  & \phantom{=} {} + \sum_{\alpha k} \left( \hat{\Phi}^\dag_{\alpha k} \vec{V}^\dagger_{\alpha k} \cdot \hat{\vec{\phi}}
  + \hat{\vec{\Phi}}^\dag \cdot \vec{V}_{\alpha k} \hat{\phi}_{\alpha k} \right) ~, \label{H_generic}
\end{align}
which is the generalization of Hamiltonian introduced in Sec.\ \ref{SEC:Hamiltonian}, Eq.\ \eqref{H} to multiple states in the molecular junction. We denote vectors in the single-particle state space of the impurity by bold symbols, e.g., $\vec{\Phi}$ or $\vec{V}_{\alpha k}$, and matrices by underlined bold symbols, e.g., $\mat{H}$. In the WBL the retarded (R) and advanced (A) Green's function are given by
\begin{subequations} \label{G_WBL}
  \begin{align}
    \mat{G}^\mathrm{R}(z) & = \sum_n \frac{\vec{R}_n \vec{A}^\dagger_n}{z - \omega_n} ~, \label{GR_WBL} \\
    \mat{G}^\mathrm{A}(z) & = \sum_m \frac{\vec{A}_m \vec{R}^\dagger_m}{z - \omega^\star_m} ~. \label{GA_WBL}
  \end{align}
\end{subequations}
The vectors $\vec{R}_n$ are the \emph{right} eigenvectors of the retarded eigenproblem and the vectors $\vec{A}_m$ are the \emph{right} eigenvectors to the corresponding advanced eigenproblem, i.e.,
\begin{subequations}
  \begin{align}
    \left( \mat{H} - i \frac{1}{2} \mat{\Gamma} \right) \cdot \vec{R}_n & = \omega_n \vec{R}_n ~, \label{Rn} \\
    \left( \mat{H} + i \frac{1}{2} \mat{\Gamma} \right) \cdot \vec{A}_n & = \omega^\star_m \vec{A}_m ~. \label{Am}
  \end{align}
\end{subequations}
They are normalized by requiring $\vec{A}_m^\dagger \cdot \vec{R}_n = \delta_{nm}$. The Landauer-B\"uttiker formula for the energy current explicitly reads
\begin{align}
  J_\alpha & = i \sum_{\alpha'}\sum_{n m} \vec{R}^\dagger_m \cdot \mat{\Gamma}_\alpha \cdot \vec{R}_n
  \vec{A}^\dagger_n \cdot \mat{\Gamma}_{\alpha'} \cdot \vec{A}_m \nonumber \\
  & \times \int_{-\infty}^{\infty}\!\!\frac{\mathrm{d}\epsilon}{2 \pi i}
  \epsilon \left[ f(\epsilon_\alpha) - f(\epsilon_{\alpha'}) \right]
  \frac{1}{\epsilon - \omega_n}\frac{1}{\epsilon - \omega^\star_m} ~, \label{Ja_WBL_1}
\end{align}
where we introduced the short hand $\epsilon_\alpha = \frac{\beta_\alpha}{\beta} (\epsilon - U_\alpha)$, with $\beta_\alpha$ being the (inverse) temperature, and $U_\alpha$ the potential in lead $\alpha$ (note that $\beta_\alpha / \beta = [1 + \psi_\alpha]^{-1}$). Using the representation of the Fermi function in terms of a Matsubara summation, i.e.,
\begin{align}
  f(z) = \frac{1}{2} - \frac{1}{\beta} \sum_f \frac{1}{z - i z_f} ~, \label{fM}
\end{align}
with the Matsubara poles $z_f = (2f+1)\pi / \beta$, we arrive at
\begin{align}
  J_\alpha & = -i \sum_{\alpha'}\sum_{n m} \vec{R}^\dagger_m \cdot \mat{\Gamma}_\alpha \cdot \vec{R}_n
  \vec{A}^\dagger_n \cdot \mat{\Gamma}_{\alpha'} \cdot \vec{A}_m \nonumber \\
  & \times \left[ C^{\alpha}_{nm} - C^{\alpha'}_{nm} \right] ~,  \label{Ja_WBL_2}
\end{align}
with 
\begin{align}
  C^{\alpha}_{nm} = \frac{1}{\beta} \sum_f 
  \int_{-\infty}^{\infty}\!\!\frac{\mathrm{d}\epsilon}{2 \pi i} 
  \frac{\epsilon}{\epsilon_\alpha - i z_f}\frac{1}{\epsilon - \omega_n}\frac{1}{\epsilon - \omega^\star_m} ~. \label{Cnm}
\end{align}
Using
\begin{align}
  \frac{\epsilon}{(\epsilon - \omega_n)(\epsilon - \omega^\star_m)}
  = \frac{\omega_n}{\omega_n - \omega^\star_m}\frac{1}{\epsilon - \omega_n}
  + \frac{\omega^\star_m}{\omega^\star_m - \omega_n}\frac{1}{\epsilon - \omega^\star_m}
\end{align}
we can further decompose
\begin{subequations}
  \begin{align}
    C^{\alpha}_{nm} & = \frac{1}{\omega_n - \omega^\star_m} \left[F^\alpha_n + (F^\alpha_m)^\star\right] ~, \label{CnmFnFm} \\
    F^\alpha_n & = \omega_n \frac{1}{\beta} \sum_f 
    \lambda^2 \int_{-\infty}^{\infty}\!\!\frac{\mathrm{d}\epsilon}{2 \pi i} 
    \frac{1}{\epsilon_\alpha - i z_f}\frac{1}{\epsilon - \omega_n} ~. \label{Fn1}
  \end{align}
\end{subequations}
The integral can be evaluated by closing the integration contour in the upper half of the complex plane, leading to
\begin{align}
  F^\alpha_n & = \omega_n \frac{1}{\beta} \sum_{f>0} 
  \frac{1}{i z_f - {\omega_n}_\alpha} ~. \label{Fn2}
\end{align}
The sum \eqref{Fn2} does not converge, but it can be combined with the corresponding sum from $C^{\alpha'}_{nm}$, i.e.,
\begin{align} 
  F^\alpha_n - F^{\alpha'}_n & = \omega_n \frac{1}{\beta} \sum_{f>0}
  \left( \frac{1}{i z_f - {\omega_n}_\alpha}
  - \frac{1}{i z_f - {\omega_n}_{\alpha'}} \right) \nonumber \\
  & = \omega_n \sum_{f>0} \frac{{\omega_n}_\alpha - {\omega_n}_{\alpha'}}
  {(i z_f - {\omega_n}_\alpha ) (i z_f - {\omega_n}_{\alpha'})} ~. \label{FnFn}
\end{align}
Expression \eqref{FnFn} can be summed explicitly using
\begin{align}
  \frac{1}{\beta} \sum_{f>0} \frac{1}{i z_f - x}\frac{1}{i z_f - y}
  = \frac{D(-x) - D(-y)}{x - y} ~, \label{FermiSum2}
\end{align}
where we defined
\begin{align}
  D(z) \equiv -\frac{1}{2 \pi i} \psi_0\left(\frac{1}{2} - i \frac{z \beta}{2 \pi}\right)
  ~, \label{D}
\end{align}
in terms of the Digamma function $\psi_0(z)$. This leads to the final result
\begin{align}
  C^{\alpha}_{nm} - C^{\alpha'}_{nm} & = 
  \frac{\omega^\star_m}{\omega^\star_m - \omega_n} \left[ D({\omega^\star_m}_\alpha) - D({\omega^\star_m}_{\alpha'}) \right] \label{CnmFinal} \\
  & + \frac{\omega_n}{\omega_n - \omega^\star_m} \left[ D(-{\omega_n}_\alpha) - D(-{\omega_n}_{\alpha'}) \right]
  ~, \nonumber
\end{align}
where we used that $[D(z)]^\star = - D(- z^\star)$, which follows from $[\psi_0(z)]^\star = \psi_0(z^\star)$, we can see that the expression for the energy current is a real number.

In the derivation presented above we have replace the frequency dependent self-energies by the frequency independent WBL approximation inside the integrand. In the following we will repeat the calculation keeping a ``minimal'' frequency dependence, i.e.,
\begin{align}
  \mat{\Sigma}^{\mathrm{R}/\mathrm{A}}_\alpha(\epsilon)
  \approx \frac{1}{2} \mat{\Gamma}_\alpha \frac{\lambda}{\epsilon \pm i \lambda} ~, \label{SWBL}
\end{align}
which reduces to the WBL as $\lambda \to \infty$. Only at the end of the calculation we will take the limit $\lambda \to \infty$. Equation \eqref{SWBL} implies that in the integral for $F^{\alpha}_{n}$ we have an additional factor of
\begin{align}
  \frac{\lambda^4}{(\epsilon-i\lambda)^2(\epsilon+i\lambda)^2}
  = \left. \partial_x \partial_y \frac{\lambda^2}
  {(\epsilon-i\lambda x)(\epsilon+i\lambda y)} \right|_{x=y=1} ~.
\end{align}
Accordingly, we have
\begin{align}
  & F^\alpha_n = \omega_n \lambda^2 \partial_{x} \partial_{y} \label{FnL1} \\
  & \Bigg(\frac{1}{\beta} \sum_{f} \frac{1}{i \lambda x_a - i z_f}
  \frac{1}{i\lambda(x+y)(i\lambda x - \omega_n)} \nonumber \\
  & + \frac{1}{\beta}\sum_{f>0} 
  \frac{\left( \frac{\beta_\alpha}{\beta} \right)^2}
  {(i z_f - i\lambda x_\alpha)(i z_f + i \lambda y_\alpha)
  (i z_f - {\omega_n}_\alpha)} \Bigg)_{x=y=1} ~. \nonumber
\end{align}
Now we use
\begin{subequations}
  \begin{align}
    & \sum_f \frac{1}{z - i z_f} = D(z) - D(-z) ~, \label{FermiSum1} \\
    & \sum_{f>0} \frac{1}{i z_f - a}\frac{1}{i z_f - c}\frac{1}{i z_f - c} \label{FermiSum3} \\
    & = \frac{D(-a)}{(a-b)(a-c)} + \frac{D(-b)}{(b-a)(b-c)}
    + \frac{D(-c)}{(c-b)(c-b)} ~, \nonumber
  \end{align}
\end{subequations}
to arrive at
\begin{align}
  F^\alpha_n & = \omega_n \lambda^2 \partial_{x} \partial_{y} 
  \Bigg( \frac{D(i\lambda x_a)}{i\lambda(x+y)(i\lambda x - \omega_n)} \nonumber \\
  & +\frac{D(i\lambda y_a)}{i\lambda(x+y)(i\lambda y - \omega_n)} \nonumber \\
  & + \frac{D(-{\omega_n}_\alpha)}{(\omega_n - i \lambda x)(\omega_n + i \lambda y)} \Bigg)_{x=y=1} ~.\label{FnL2}
\end{align}
From the asymptotic expansion of the Digamma function, $\psi_0(z) \sim \log(z)$, it follows that
\begin{align}
  D(i\lambda z_\alpha) \sim -\frac{\log\left[\frac{z\beta_\alpha}{2 \pi}\right]}{2 \pi i}
  ~, \label{Dasymptotic}
\end{align}
which, in turn, leads to the asymptotic expansion 
\begin{align}
  F^\alpha_n & \sim \omega_n \partial_x \partial_y 
  \Bigg( \frac{D(-{\omega_n}_\alpha)}{xy} \nonumber \\
  & + \frac{\log\left[\frac{\lambda x \beta_\alpha}{2 \pi}\right]}{2 \pi i(x+y)x} 
  + \frac{\log\left[\frac{\lambda y \beta_\alpha}{2 \pi}\right]}{2 \pi i(x+y)y}
  \Bigg)_{x=y=1} ~. \label{FnL3}
\end{align}
Combining the terms due to the different leads yields
\begin{align}
  F^\alpha_n - F^{\alpha'}_n & \sim \omega_n \bigg(D(-{\omega_n}_\alpha)
  - D(-{\omega_n}_{\alpha'}) \nonumber \\
  & \phantom{\sim \omega_n \bigg(} {} - i \frac{1}{2\pi}
  \log\left[\beta_\alpha / \beta_{\alpha'}\right] \bigg) ~. \label{FnaFna}
\end{align}
Equation \eqref{FnaFna} does not depend on $\lambda$ anymore and we can safely take the limit $\lambda \to \infty$, because the neglected terms in the asymptotic expansion are of order $\lambda^{-1}$. Plugging Eq.\ \eqref{FnaFna} into Eqs.\ \eqref{CnmFnFm} and \eqref{Ja_WBL_2} finally gives
\begin{align}
  J_\alpha & = -i \sum_{\alpha'}\sum_{n m} \vec{R}^\dagger_m \cdot \mat{\Gamma}_\alpha \cdot \vec{R}_n
  \vec{A}^\dagger_n \cdot \mat{\Gamma}_{\alpha'} \cdot \vec{A}_m \nonumber \\
  & \times \Bigg( \frac{\omega_n}{\omega_n - \omega^\star_m} 
  \left[D(-{\omega_n}_\alpha) - D(-{\omega_n}_{\alpha'})\right] \nonumber \\
  & \phantom{\times \Bigg(} {} + \frac{\omega^\star_m}{\omega^\star_m - \omega_n} 
    \left[D({\omega^\star_m}_\alpha) - D({\omega^\star_m}_{\alpha'})\right] \Bigg) \nonumber \\
  & + \frac{1}{2 \pi} \sum_{\alpha'} \mathrm{Tr} \left[ \mat{\Gamma}_\alpha
  \cdot \mat{\Gamma}_{\alpha'} \right]
  \log\left(\frac{T_\alpha }{T_{\alpha'}} \right) ~. \label{Ja_WBL}
\end{align}
The first term corresponds to the result when taking the WBL inside the integral. The second term is the correction if, instead, the WBL is taken after the integration. This correction term vanishes if there is no temperature gradient between the leads. Furthermore, it vanishes if the coupling matrices do not overlap in the single-particle state space of the molecular Hamiltonian. In the results presented in this work this correction is crucial in order to reproduce the steady-state heat currents in the WBL.

\end{appendix}

\bibliography{ThermoWBL}

\begin{thebibliography}{46}%
\makeatletter
\providecommand \@ifxundefined [1]{%
 \@ifx{#1\undefined}
}%
\providecommand \@ifnum [1]{%
 \ifnum #1\expandafter \@firstoftwo
 \else \expandafter \@secondoftwo
 \fi
}%
\providecommand \@ifx [1]{%
 \ifx #1\expandafter \@firstoftwo
 \else \expandafter \@secondoftwo
 \fi
}%
\providecommand \natexlab [1]{#1}%
\providecommand \enquote  [1]{``#1''}%
\providecommand \bibnamefont  [1]{#1}%
\providecommand \bibfnamefont [1]{#1}%
\providecommand \citenamefont [1]{#1}%
\providecommand \href@noop [0]{\@secondoftwo}%
\providecommand \href [0]{\begingroup \@sanitize@url \@href}%
\providecommand \@href[1]{\@@startlink{#1}\@@href}%
\providecommand \@@href[1]{\endgroup#1\@@endlink}%
\providecommand \@sanitize@url [0]{\catcode `\\12\catcode `\$12\catcode
  `\&12\catcode `\#12\catcode `\^12\catcode `\_12\catcode `\%12\relax}%
\providecommand \@@startlink[1]{}%
\providecommand \@@endlink[0]{}%
\providecommand \url  [0]{\begingroup\@sanitize@url \@url }%
\providecommand \@url [1]{\endgroup\@href {#1}{\urlprefix }}%
\providecommand \urlprefix  [0]{URL }%
\providecommand \Eprint [0]{\href }%
\providecommand \doibase [0]{http://dx.doi.org/}%
\providecommand \selectlanguage [0]{\@gobble}%
\providecommand \bibinfo  [0]{\@secondoftwo}%
\providecommand \bibfield  [0]{\@secondoftwo}%
\providecommand \translation [1]{[#1]}%
\providecommand \BibitemOpen [0]{}%
\providecommand \bibitemStop [0]{}%
\providecommand \bibitemNoStop [0]{.\EOS\space}%
\providecommand \EOS [0]{\spacefactor3000\relax}%
\providecommand \BibitemShut  [1]{\csname bibitem#1\endcsname}%
\let\auto@bib@innerbib\@empty
%</preamble>
\bibitem [{\citenamefont {Landauer}(1957)}]{Landauer:57}%
  \BibitemOpen
  \bibfield  {author} {\bibinfo {author} {\bibfnamefont {R.}~\bibnamefont
  {Landauer}},\ }\href {\doibase 10.1147/rd.13.0223} {\bibfield  {journal}
  {\bibinfo  {journal} {IBM J. Research and Development}\ }\textbf {\bibinfo
  {volume} {1}},\ \bibinfo {pages} {223} (\bibinfo {year} {1957})}\BibitemShut
  {NoStop}%
\bibitem [{\citenamefont {B{\"u}ttiker}\ \emph {et~al.}(1985)\citenamefont
  {B{\"u}ttiker}, \citenamefont {Imry}, \citenamefont {Landauer},\ and\
  \citenamefont {Pinhas}}]{BuettikerPinhas:85}%
  \BibitemOpen
  \bibfield  {author} {\bibinfo {author} {\bibfnamefont {M.}~\bibnamefont
  {B{\"u}ttiker}}, \bibinfo {author} {\bibfnamefont {Y.}~\bibnamefont {Imry}},
  \bibinfo {author} {\bibfnamefont {R.}~\bibnamefont {Landauer}}, \ and\
  \bibinfo {author} {\bibfnamefont {S.}~\bibnamefont {Pinhas}},\ }\href
  {\doibase 10.1103/PhysRevB.31.6207} {\bibfield  {journal} {\bibinfo
  {journal} {Phys. Rev. B}\ }\textbf {\bibinfo {volume} {31}},\ \bibinfo
  {pages} {6207} (\bibinfo {year} {1985})}\BibitemShut {NoStop}%
\bibitem [{\citenamefont {Landauer}(1989)}]{Landauer:89}%
  \BibitemOpen
  \bibfield  {author} {\bibinfo {author} {\bibfnamefont {R.}~\bibnamefont
  {Landauer}},\ }\href {http://stacks.iop.org/0953-8984/1/i=43/a=011}
  {\bibfield  {journal} {\bibinfo  {journal} {J. Phys.: Condens. Matter}\
  }\textbf {\bibinfo {volume} {1}},\ \bibinfo {pages} {8099} (\bibinfo {year}
  {1989})}\BibitemShut {NoStop}%
\bibitem [{\citenamefont {Goldsmid}(2009)}]{Goldsmid:09}%
  \BibitemOpen
  \bibfield  {author} {\bibinfo {author} {\bibfnamefont {H.~J.}\ \bibnamefont
  {Goldsmid}},\ }\href@noop {} {\emph {\bibinfo {title} {Introduction to
  Thermoelectricity}}},\ Springer Series in Materials Science\ (\bibinfo
  {publisher} {Springer},\ \bibinfo {year} {2009})\BibitemShut {NoStop}%
\bibitem [{\citenamefont {Dubi}\ and\ \citenamefont
  {Di~Ventra}(2009)}]{DubiDiVentra:09}%
  \BibitemOpen
  \bibfield  {author} {\bibinfo {author} {\bibfnamefont {Y.}~\bibnamefont
  {Dubi}}\ and\ \bibinfo {author} {\bibfnamefont {M.}~\bibnamefont
  {Di~Ventra}},\ }\href {\doibase 10.1021/nl8025407} {\bibfield  {journal}
  {\bibinfo  {journal} {Nano Lett.}\ }\textbf {\bibinfo {volume} {9}},\
  \bibinfo {pages} {97} (\bibinfo {year} {2009})},\ \Eprint
  {http://arxiv.org/abs/http://pubs.acs.org/doi/pdf/10.1021/nl8025407}
  {http://pubs.acs.org/doi/pdf/10.1021/nl8025407} \BibitemShut {NoStop}%
\bibitem [{\citenamefont {Mecklenburg}\ \emph {et~al.}(2015)\citenamefont
  {Mecklenburg}, \citenamefont {Hubbard}, \citenamefont {White}, \citenamefont
  {Dhall}, \citenamefont {Cronin}, \citenamefont {Aloni},\ and\ \citenamefont
  {Regan}}]{MecklenburgRegan:15}%
  \BibitemOpen
  \bibfield  {author} {\bibinfo {author} {\bibfnamefont {M.}~\bibnamefont
  {Mecklenburg}}, \bibinfo {author} {\bibfnamefont {W.~A.}\ \bibnamefont
  {Hubbard}}, \bibinfo {author} {\bibfnamefont {E.~R.}\ \bibnamefont {White}},
  \bibinfo {author} {\bibfnamefont {R.}~\bibnamefont {Dhall}}, \bibinfo
  {author} {\bibfnamefont {S.~B.}\ \bibnamefont {Cronin}}, \bibinfo {author}
  {\bibfnamefont {S.}~\bibnamefont {Aloni}}, \ and\ \bibinfo {author}
  {\bibfnamefont {B.~C.}\ \bibnamefont {Regan}},\ }\href {\doibase
  10.1126/science.aaa2433} {\bibfield  {journal} {\bibinfo  {journal}
  {Science}\ }\textbf {\bibinfo {volume} {347}},\ \bibinfo {pages} {629}
  (\bibinfo {year} {2015})},\ \Eprint
  {http://arxiv.org/abs/http://www.sciencemag.org/content/347/6222/629.full.pdf}
  {http://www.sciencemag.org/content/347/6222/629.full.pdf} \BibitemShut
  {NoStop}%
\bibitem [{\citenamefont {Halbertal}\ \emph {et~al.}(2016)\citenamefont
  {Halbertal}, \citenamefont {Cuppens}, \citenamefont {Shalom}, \citenamefont
  {Embon}, \citenamefont {Shadmi}, \citenamefont {Anahory}, \citenamefont
  {Naren}, \citenamefont {Sarkar}, \citenamefont {Uri}, \citenamefont {Ronen},
  \citenamefont {Myasoedov}, \citenamefont {Levitov}, \citenamefont
  {Joselevich}, \citenamefont {Geim},\ and\ \citenamefont
  {Zeldov}}]{HalbertalZeldov:16}%
  \BibitemOpen
  \bibfield  {author} {\bibinfo {author} {\bibfnamefont {D.}~\bibnamefont
  {Halbertal}}, \bibinfo {author} {\bibfnamefont {J.}~\bibnamefont {Cuppens}},
  \bibinfo {author} {\bibfnamefont {M.~B.}\ \bibnamefont {Shalom}}, \bibinfo
  {author} {\bibfnamefont {L.}~\bibnamefont {Embon}}, \bibinfo {author}
  {\bibfnamefont {N.}~\bibnamefont {Shadmi}}, \bibinfo {author} {\bibfnamefont
  {Y.}~\bibnamefont {Anahory}}, \bibinfo {author} {\bibfnamefont {H.~R.}\
  \bibnamefont {Naren}}, \bibinfo {author} {\bibfnamefont {J.}~\bibnamefont
  {Sarkar}}, \bibinfo {author} {\bibfnamefont {A.}~\bibnamefont {Uri}},
  \bibinfo {author} {\bibfnamefont {Y.}~\bibnamefont {Ronen}}, \bibinfo
  {author} {\bibfnamefont {Y.}~\bibnamefont {Myasoedov}}, \bibinfo {author}
  {\bibfnamefont {L.~S.}\ \bibnamefont {Levitov}}, \bibinfo {author}
  {\bibfnamefont {E.}~\bibnamefont {Joselevich}}, \bibinfo {author}
  {\bibfnamefont {A.~K.}\ \bibnamefont {Geim}}, \ and\ \bibinfo {author}
  {\bibfnamefont {E.}~\bibnamefont {Zeldov}},\ }\href
  {http://dx.doi.org/10.1038/nature19843} {\bibfield  {journal} {\bibinfo
  {journal} {Nature}\ }\textbf {\bibinfo {volume} {539}},\ \bibinfo {pages}
  {407} (\bibinfo {year} {2016})},\ \bibinfo {note} {letter}\BibitemShut
  {NoStop}%
\bibitem [{\citenamefont {Kinoshita}\ \emph {et~al.}(1995)\citenamefont
  {Kinoshita}, \citenamefont {Misu},\ and\ \citenamefont
  {Munakata}}]{KinoshitaMunakata:95}%
  \BibitemOpen
  \bibfield  {author} {\bibinfo {author} {\bibfnamefont {I.}~\bibnamefont
  {Kinoshita}}, \bibinfo {author} {\bibfnamefont {A.}~\bibnamefont {Misu}}, \
  and\ \bibinfo {author} {\bibfnamefont {T.}~\bibnamefont {Munakata}},\ }\href
  {\doibase 10.1063/1.468605} {\bibfield  {journal} {\bibinfo  {journal} {The
  Journal of Chemical Physics}\ }\textbf {\bibinfo {volume} {102}},\ \bibinfo
  {pages} {2970} (\bibinfo {year} {1995})}\BibitemShut {NoStop}%
\bibitem [{\citenamefont {Gauyacq}\ \emph {et~al.}(2001)\citenamefont
  {Gauyacq}, \citenamefont {Borisov},\ and\ \citenamefont
  {Raşeev}}]{GauyacqRaseev:01}%
  \BibitemOpen
  \bibfield  {author} {\bibinfo {author} {\bibfnamefont {J.}~\bibnamefont
  {Gauyacq}}, \bibinfo {author} {\bibfnamefont {A.}~\bibnamefont {Borisov}}, \
  and\ \bibinfo {author} {\bibfnamefont {G.}~\bibnamefont {Raşeev}},\ }\href
  {\doibase https://doi.org/10.1016/S0039-6028(01)01229-8} {\bibfield
  {journal} {\bibinfo  {journal} {Surface Science}\ }\textbf {\bibinfo {volume}
  {490}},\ \bibinfo {pages} {99 } (\bibinfo {year} {2001})}\BibitemShut
  {NoStop}%
\bibitem [{\citenamefont {Kirchmann}\ \emph {et~al.}(2005)\citenamefont
  {Kirchmann}, \citenamefont {Loukakos}, \citenamefont {Bovensiepen},\ and\
  \citenamefont {Wolf}}]{KirchmannWolf:05}%
  \BibitemOpen
  \bibfield  {author} {\bibinfo {author} {\bibfnamefont {P.~S.}\ \bibnamefont
  {Kirchmann}}, \bibinfo {author} {\bibfnamefont {P.~A.}\ \bibnamefont
  {Loukakos}}, \bibinfo {author} {\bibfnamefont {U.}~\bibnamefont
  {Bovensiepen}}, \ and\ \bibinfo {author} {\bibfnamefont {M.}~\bibnamefont
  {Wolf}},\ }\href {http://stacks.iop.org/1367-2630/7/i=1/a=113} {\bibfield
  {journal} {\bibinfo  {journal} {New Journal of Physics}\ }\textbf {\bibinfo
  {volume} {7}},\ \bibinfo {pages} {113} (\bibinfo {year} {2005})}\BibitemShut
  {NoStop}%
\bibitem [{\citenamefont {Chulkov}\ \emph {et~al.}(2006)\citenamefont
  {Chulkov}, \citenamefont {Borisov}, \citenamefont {Gauyacq}, \citenamefont
  {S{\'a}nchez-Portal}, \citenamefont {Silkin}, \citenamefont {Zhukov},\ and\
  \citenamefont {Echenique}}]{ChulkovEchenique:06}%
  \BibitemOpen
  \bibfield  {author} {\bibinfo {author} {\bibfnamefont {E.~V.}\ \bibnamefont
  {Chulkov}}, \bibinfo {author} {\bibfnamefont {A.~G.}\ \bibnamefont
  {Borisov}}, \bibinfo {author} {\bibfnamefont {J.~P.}\ \bibnamefont
  {Gauyacq}}, \bibinfo {author} {\bibfnamefont {D.}~\bibnamefont
  {S{\'a}nchez-Portal}}, \bibinfo {author} {\bibfnamefont {V.~M.}\ \bibnamefont
  {Silkin}}, \bibinfo {author} {\bibfnamefont {V.~P.}\ \bibnamefont {Zhukov}},
  \ and\ \bibinfo {author} {\bibfnamefont {P.~M.}\ \bibnamefont {Echenique}},\
  }\href {\doibase 10.1021/cr050166o} {\bibfield  {journal} {\bibinfo
  {journal} {Chemical Reviews}\ }\textbf {\bibinfo {volume} {106}},\ \bibinfo
  {pages} {4160} (\bibinfo {year} {2006})}\BibitemShut {NoStop}%
\bibitem [{\citenamefont {Myllyperki{\"o}}\ \emph {et~al.}(2010)\citenamefont
  {Myllyperki{\"o}}, \citenamefont {Herranen}, \citenamefont {Rintala},
  \citenamefont {Jiang}, \citenamefont {Mudimela}, \citenamefont {Zhu},
  \citenamefont {Nasibulin}, \citenamefont {Johansson}, \citenamefont
  {Kauppinen}, \citenamefont {Ahlskog},\ and\ \citenamefont
  {Pettersson}}]{MyllyperkioPettersson:10}%
  \BibitemOpen
  \bibfield  {author} {\bibinfo {author} {\bibfnamefont {P.}~\bibnamefont
  {Myllyperki{\"o}}}, \bibinfo {author} {\bibfnamefont {O.}~\bibnamefont
  {Herranen}}, \bibinfo {author} {\bibfnamefont {J.}~\bibnamefont {Rintala}},
  \bibinfo {author} {\bibfnamefont {H.}~\bibnamefont {Jiang}}, \bibinfo
  {author} {\bibfnamefont {P.~R.}\ \bibnamefont {Mudimela}}, \bibinfo {author}
  {\bibfnamefont {Z.}~\bibnamefont {Zhu}}, \bibinfo {author} {\bibfnamefont
  {A.~G.}\ \bibnamefont {Nasibulin}}, \bibinfo {author} {\bibfnamefont
  {A.}~\bibnamefont {Johansson}}, \bibinfo {author} {\bibfnamefont {E.~I.}\
  \bibnamefont {Kauppinen}}, \bibinfo {author} {\bibfnamefont {M.}~\bibnamefont
  {Ahlskog}}, \ and\ \bibinfo {author} {\bibfnamefont {M.}~\bibnamefont
  {Pettersson}},\ }\href {\doibase 10.1021/nn1015067} {\bibfield  {journal}
  {\bibinfo  {journal} {ACS Nano}\ }\textbf {\bibinfo {volume} {4}},\ \bibinfo
  {pages} {6780} (\bibinfo {year} {2010})}\BibitemShut {NoStop}%
\bibitem [{\citenamefont {Cocker}\ \emph {et~al.}(2013)\citenamefont {Cocker},
  \citenamefont {Jelic}, \citenamefont {Gupta}, \citenamefont {Molesky},
  \citenamefont {Burgess}, \citenamefont {Reyes}, \citenamefont {Titova},
  \citenamefont {Tsui}, \citenamefont {Freeman},\ and\ \citenamefont
  {Hegmann}}]{CockerHegmann:13}%
  \BibitemOpen
  \bibfield  {author} {\bibinfo {author} {\bibfnamefont {T.~L.}\ \bibnamefont
  {Cocker}}, \bibinfo {author} {\bibfnamefont {V.}~\bibnamefont {Jelic}},
  \bibinfo {author} {\bibfnamefont {M.}~\bibnamefont {Gupta}}, \bibinfo
  {author} {\bibfnamefont {S.~J.}\ \bibnamefont {Molesky}}, \bibinfo {author}
  {\bibfnamefont {J.~A.~J.}\ \bibnamefont {Burgess}}, \bibinfo {author}
  {\bibfnamefont {G.~D.~L.}\ \bibnamefont {Reyes}}, \bibinfo {author}
  {\bibfnamefont {L.~V.}\ \bibnamefont {Titova}}, \bibinfo {author}
  {\bibfnamefont {Y.~Y.}\ \bibnamefont {Tsui}}, \bibinfo {author}
  {\bibfnamefont {M.~R.}\ \bibnamefont {Freeman}}, \ and\ \bibinfo {author}
  {\bibfnamefont {F.~A.}\ \bibnamefont {Hegmann}},\ }\href
  {http://dx.doi.org/10.1038/nphoton.2013.151} {\bibfield  {journal} {\bibinfo
  {journal} {Nature Photonics}\ }\textbf {\bibinfo {volume} {7}},\ \bibinfo
  {pages} {620} (\bibinfo {year} {2013})}\BibitemShut {NoStop}%
\bibitem [{\citenamefont {Ni}\ \emph {et~al.}(2016)\citenamefont {Ni},
  \citenamefont {Wang}, \citenamefont {Goldflam}, \citenamefont {Wagner},
  \citenamefont {Fei}, \citenamefont {McLeod}, \citenamefont {Liu},
  \citenamefont {Keilmann}, \citenamefont {{\"O}zyilmaz}, \citenamefont
  {Castro~Neto}, \citenamefont {Hone}, \citenamefont {Fogler},\ and\
  \citenamefont {Basov}}]{NiBasov:16}%
  \BibitemOpen
  \bibfield  {author} {\bibinfo {author} {\bibfnamefont {G.~X.}\ \bibnamefont
  {Ni}}, \bibinfo {author} {\bibfnamefont {L.}~\bibnamefont {Wang}}, \bibinfo
  {author} {\bibfnamefont {M.~D.}\ \bibnamefont {Goldflam}}, \bibinfo {author}
  {\bibfnamefont {M.}~\bibnamefont {Wagner}}, \bibinfo {author} {\bibfnamefont
  {Z.}~\bibnamefont {Fei}}, \bibinfo {author} {\bibfnamefont {A.~S.}\
  \bibnamefont {McLeod}}, \bibinfo {author} {\bibfnamefont {M.~K.}\
  \bibnamefont {Liu}}, \bibinfo {author} {\bibfnamefont {F.}~\bibnamefont
  {Keilmann}}, \bibinfo {author} {\bibfnamefont {B.}~\bibnamefont
  {{\"O}zyilmaz}}, \bibinfo {author} {\bibfnamefont {A.~H.}\ \bibnamefont
  {Castro~Neto}}, \bibinfo {author} {\bibfnamefont {J.}~\bibnamefont {Hone}},
  \bibinfo {author} {\bibfnamefont {M.~M.}\ \bibnamefont {Fogler}}, \ and\
  \bibinfo {author} {\bibfnamefont {D.~N.}\ \bibnamefont {Basov}},\ }\href
  {http://dx.doi.org/10.1038/nphoton.2016.45} {\bibfield  {journal} {\bibinfo
  {journal} {Nature Photonics}\ }\textbf {\bibinfo {volume} {10}},\ \bibinfo
  {pages} {244} (\bibinfo {year} {2016})}\BibitemShut {NoStop}%
\bibitem [{\citenamefont {{Karnetzky}}\ \emph {et~al.}(2017)\citenamefont
  {{Karnetzky}}, \citenamefont {{Zimmermann}}, \citenamefont {{Trummer}},
  \citenamefont {{Duque-Sierra}}, \citenamefont {{W{\"o}rle}}, \citenamefont
  {{Kienberger}},\ and\ \citenamefont
  {{Holleitner}}}]{KarnetzkyHolleitner:17arxiv}%
  \BibitemOpen
  \bibfield  {author} {\bibinfo {author} {\bibfnamefont {C.}~\bibnamefont
  {{Karnetzky}}}, \bibinfo {author} {\bibfnamefont {P.}~\bibnamefont
  {{Zimmermann}}}, \bibinfo {author} {\bibfnamefont {C.}~\bibnamefont
  {{Trummer}}}, \bibinfo {author} {\bibfnamefont {C.}~\bibnamefont
  {{Duque-Sierra}}}, \bibinfo {author} {\bibfnamefont {M.}~\bibnamefont
  {{W{\"o}rle}}}, \bibinfo {author} {\bibfnamefont {R.}~\bibnamefont
  {{Kienberger}}}, \ and\ \bibinfo {author} {\bibfnamefont {A.}~\bibnamefont
  {{Holleitner}}},\ }\href@noop {} {\bibfield  {journal} {\bibinfo  {journal}
  {ArXiv e-prints}\ } (\bibinfo {year} {2017})},\ \Eprint
  {http://arxiv.org/abs/1708.00262} {arXiv:1708.00262 [cond-mat.mes-hall]}
  \BibitemShut {NoStop}%
\bibitem [{\citenamefont {Cini}(1980)}]{Cini:80}%
  \BibitemOpen
  \bibfield  {author} {\bibinfo {author} {\bibfnamefont {M.}~\bibnamefont
  {Cini}},\ }\href {\doibase 10.1103/PhysRevB.22.5887} {\bibfield  {journal}
  {\bibinfo  {journal} {Phys. Rev. B}\ }\textbf {\bibinfo {volume} {22}},\
  \bibinfo {pages} {5887} (\bibinfo {year} {1980})}\BibitemShut {NoStop}%
\bibitem [{\citenamefont {Stefanucci}\ and\ \citenamefont
  {Almbladh}(2004)}]{StefanucciAlmbladh:04}%
  \BibitemOpen
  \bibfield  {author} {\bibinfo {author} {\bibfnamefont {G.}~\bibnamefont
  {Stefanucci}}\ and\ \bibinfo {author} {\bibfnamefont {C.-O.}\ \bibnamefont
  {Almbladh}},\ }\href {\doibase 10.1103/PhysRevB.69.195318} {\bibfield
  {journal} {\bibinfo  {journal} {Phys. Rev. B}\ }\textbf {\bibinfo {volume}
  {69}},\ \bibinfo {pages} {195318} (\bibinfo {year} {2004})}\BibitemShut
  {NoStop}%
\bibitem [{\citenamefont {Luttinger}(1964)}]{Luttinger:64a}%
  \BibitemOpen
  \bibfield  {author} {\bibinfo {author} {\bibfnamefont {J.~M.}\ \bibnamefont
  {Luttinger}},\ }\href {\doibase 10.1103/PhysRev.135.A1505} {\bibfield
  {journal} {\bibinfo  {journal} {Phys. Rev.}\ }\textbf {\bibinfo {volume}
  {135}},\ \bibinfo {pages} {A1505} (\bibinfo {year} {1964})}\BibitemShut
  {NoStop}%
\bibitem [{\citenamefont {Shastry}(2009)}]{Shastry:09}%
  \BibitemOpen
  \bibfield  {author} {\bibinfo {author} {\bibfnamefont {B.~S.}\ \bibnamefont
  {Shastry}},\ }\href {http://stacks.iop.org/0034-4885/72/i=1/a=016501}
  {\bibfield  {journal} {\bibinfo  {journal} {Rep. Prog. Phys.}\ }\textbf
  {\bibinfo {volume} {72}},\ \bibinfo {pages} {016501} (\bibinfo {year}
  {2009})}\BibitemShut {NoStop}%
\bibitem [{\citenamefont {Eich}\ \emph
  {et~al.}(2014{\natexlab{a}})\citenamefont {Eich}, \citenamefont {Principi},
  \citenamefont {Di~Ventra},\ and\ \citenamefont {Vignale}}]{EichVignale:14b}%
  \BibitemOpen
  \bibfield  {author} {\bibinfo {author} {\bibfnamefont {F.~G.}\ \bibnamefont
  {Eich}}, \bibinfo {author} {\bibfnamefont {A.}~\bibnamefont {Principi}},
  \bibinfo {author} {\bibfnamefont {M.}~\bibnamefont {Di~Ventra}}, \ and\
  \bibinfo {author} {\bibfnamefont {G.}~\bibnamefont {Vignale}},\ }\href
  {\doibase 10.1103/PhysRevB.90.115116} {\bibfield  {journal} {\bibinfo
  {journal} {Phys. Rev. B}\ }\textbf {\bibinfo {volume} {90}},\ \bibinfo
  {pages} {115116} (\bibinfo {year} {2014}{\natexlab{a}})}\BibitemShut
  {NoStop}%
\bibitem [{\citenamefont {Zhu}\ \emph {et~al.}(2005)\citenamefont {Zhu},
  \citenamefont {Maciejko}, \citenamefont {Ji}, \citenamefont {Guo},\ and\
  \citenamefont {Wang}}]{ZhuWang:05}%
  \BibitemOpen
  \bibfield  {author} {\bibinfo {author} {\bibfnamefont {Y.}~\bibnamefont
  {Zhu}}, \bibinfo {author} {\bibfnamefont {J.}~\bibnamefont {Maciejko}},
  \bibinfo {author} {\bibfnamefont {T.}~\bibnamefont {Ji}}, \bibinfo {author}
  {\bibfnamefont {H.}~\bibnamefont {Guo}}, \ and\ \bibinfo {author}
  {\bibfnamefont {J.}~\bibnamefont {Wang}},\ }\href {\doibase
  10.1103/PhysRevB.71.075317} {\bibfield  {journal} {\bibinfo  {journal} {Phys.
  Rev. B}\ }\textbf {\bibinfo {volume} {71}},\ \bibinfo {pages} {075317}
  (\bibinfo {year} {2005})}\BibitemShut {NoStop}%
\bibitem [{\citenamefont {Maciejko}\ \emph {et~al.}(2006)\citenamefont
  {Maciejko}, \citenamefont {Wang},\ and\ \citenamefont
  {Guo}}]{MaciejkoGuo:06}%
  \BibitemOpen
  \bibfield  {author} {\bibinfo {author} {\bibfnamefont {J.}~\bibnamefont
  {Maciejko}}, \bibinfo {author} {\bibfnamefont {J.}~\bibnamefont {Wang}}, \
  and\ \bibinfo {author} {\bibfnamefont {H.}~\bibnamefont {Guo}},\ }\href
  {\doibase 10.1103/PhysRevB.74.085324} {\bibfield  {journal} {\bibinfo
  {journal} {Phys. Rev. B}\ }\textbf {\bibinfo {volume} {74}},\ \bibinfo
  {pages} {085324} (\bibinfo {year} {2006})}\BibitemShut {NoStop}%
\bibitem [{\citenamefont {Zheng}\ \emph {et~al.}(2007)\citenamefont {Zheng},
  \citenamefont {Wang}, \citenamefont {Yam}, \citenamefont {Mo},\ and\
  \citenamefont {Chen}}]{ZhengChen:07}%
  \BibitemOpen
  \bibfield  {author} {\bibinfo {author} {\bibfnamefont {X.}~\bibnamefont
  {Zheng}}, \bibinfo {author} {\bibfnamefont {F.}~\bibnamefont {Wang}},
  \bibinfo {author} {\bibfnamefont {C.~Y.}\ \bibnamefont {Yam}}, \bibinfo
  {author} {\bibfnamefont {Y.}~\bibnamefont {Mo}}, \ and\ \bibinfo {author}
  {\bibfnamefont {G.}~\bibnamefont {Chen}},\ }\href {\doibase
  10.1103/PhysRevB.75.195127} {\bibfield  {journal} {\bibinfo  {journal} {Phys.
  Rev. B}\ }\textbf {\bibinfo {volume} {75}},\ \bibinfo {pages} {195127}
  (\bibinfo {year} {2007})}\BibitemShut {NoStop}%
\bibitem [{\citenamefont {Zhang}\ \emph {et~al.}(2013)\citenamefont {Zhang},
  \citenamefont {Chen},\ and\ \citenamefont {Chen}}]{ZhangChen:13}%
  \BibitemOpen
  \bibfield  {author} {\bibinfo {author} {\bibfnamefont {Y.}~\bibnamefont
  {Zhang}}, \bibinfo {author} {\bibfnamefont {S.}~\bibnamefont {Chen}}, \ and\
  \bibinfo {author} {\bibfnamefont {G.}~\bibnamefont {Chen}},\ }\href {\doibase
  10.1103/PhysRevB.87.085110} {\bibfield  {journal} {\bibinfo  {journal} {Phys.
  Rev. B}\ }\textbf {\bibinfo {volume} {87}},\ \bibinfo {pages} {085110}
  (\bibinfo {year} {2013})}\BibitemShut {NoStop}%
\bibitem [{\citenamefont {Verzijl}\ \emph {et~al.}(2013)\citenamefont
  {Verzijl}, \citenamefont {Seldenthuis},\ and\ \citenamefont
  {Thijssen}}]{VerzijlThijssen:13}%
  \BibitemOpen
  \bibfield  {author} {\bibinfo {author} {\bibfnamefont {C.~J.~O.}\
  \bibnamefont {Verzijl}}, \bibinfo {author} {\bibfnamefont {J.~S.}\
  \bibnamefont {Seldenthuis}}, \ and\ \bibinfo {author} {\bibfnamefont {J.~M.}\
  \bibnamefont {Thijssen}},\ }\href {\doibase 10.1063/1.4793259} {\bibfield
  {journal} {\bibinfo  {journal} {The Journal of Chemical Physics}\ }\textbf
  {\bibinfo {volume} {138}},\ \bibinfo {pages} {094102} (\bibinfo {year}
  {2013})}\BibitemShut {NoStop}%
\bibitem [{\citenamefont {Shi}\ \emph {et~al.}(2016)\citenamefont {Shi},
  \citenamefont {Hu}, \citenamefont {Ying},\ and\ \citenamefont
  {Jin}}]{ShiJin:16}%
  \BibitemOpen
  \bibfield  {author} {\bibinfo {author} {\bibfnamefont {P.}~\bibnamefont
  {Shi}}, \bibinfo {author} {\bibfnamefont {M.}~\bibnamefont {Hu}}, \bibinfo
  {author} {\bibfnamefont {Y.}~\bibnamefont {Ying}}, \ and\ \bibinfo {author}
  {\bibfnamefont {J.}~\bibnamefont {Jin}},\ }\href {\doibase 10.1063/1.4962527}
  {\bibfield  {journal} {\bibinfo  {journal} {AIP Advances}\ }\textbf {\bibinfo
  {volume} {6}},\ \bibinfo {pages} {095002} (\bibinfo {year}
  {2016})}\BibitemShut {NoStop}%
\bibitem [{\citenamefont {B{\^{a}}ldea}(2016)}]{Baldea:16}%
  \BibitemOpen
  \bibfield  {author} {\bibinfo {author} {\bibfnamefont {I.}~\bibnamefont
  {B{\^{a}}ldea}},\ }\href {\doibase 10.3762/bjnano.7.37} {\bibfield  {journal}
  {\bibinfo  {journal} {Beilstein Journal of Nanotechnology}\ }\textbf
  {\bibinfo {volume} {7}},\ \bibinfo {pages} {418} (\bibinfo {year}
  {2016})}\BibitemShut {NoStop}%
\bibitem [{Note1()}]{Note1}%
  \BibitemOpen
  \bibinfo {note} {The function $S(z)$ has a branch cut on the real axis from
  $z=-1 \to z=1$.}\BibitemShut {Stop}%
\bibitem [{\citenamefont {Arrachea}\ \emph {et~al.}(2007)\citenamefont
  {Arrachea}, \citenamefont {Moskalets},\ and\ \citenamefont
  {Martin-Moreno}}]{ArracheaMartinMoreno:07}%
  \BibitemOpen
  \bibfield  {author} {\bibinfo {author} {\bibfnamefont {L.}~\bibnamefont
  {Arrachea}}, \bibinfo {author} {\bibfnamefont {M.}~\bibnamefont {Moskalets}},
  \ and\ \bibinfo {author} {\bibfnamefont {L.}~\bibnamefont {Martin-Moreno}},\
  }\href {\doibase 10.1103/PhysRevB.75.245420} {\bibfield  {journal} {\bibinfo
  {journal} {Phys. Rev. B}\ }\textbf {\bibinfo {volume} {75}},\ \bibinfo
  {pages} {245420} (\bibinfo {year} {2007})}\BibitemShut {NoStop}%
\bibitem [{\citenamefont {Esposito}\ \emph {et~al.}(2015)\citenamefont
  {Esposito}, \citenamefont {Ochoa},\ and\ \citenamefont
  {Galperin}}]{EspositoGalperin:15b}%
  \BibitemOpen
  \bibfield  {author} {\bibinfo {author} {\bibfnamefont {M.}~\bibnamefont
  {Esposito}}, \bibinfo {author} {\bibfnamefont {M.~A.}\ \bibnamefont {Ochoa}},
  \ and\ \bibinfo {author} {\bibfnamefont {M.}~\bibnamefont {Galperin}},\
  }\href {\doibase 10.1103/PhysRevB.92.235440} {\bibfield  {journal} {\bibinfo
  {journal} {Phys. Rev. B}\ }\textbf {\bibinfo {volume} {92}},\ \bibinfo
  {pages} {235440} (\bibinfo {year} {2015})}\BibitemShut {NoStop}%
\bibitem [{\citenamefont {Ludovico}\ \emph {et~al.}(2016)\citenamefont
  {Ludovico}, \citenamefont {Arrachea}, \citenamefont {Moskalets},\ and\
  \citenamefont {S{\'{a}}nchez}}]{LudovicoSanchez:16}%
  \BibitemOpen
  \bibfield  {author} {\bibinfo {author} {\bibfnamefont {M.}~\bibnamefont
  {Ludovico}}, \bibinfo {author} {\bibfnamefont {L.}~\bibnamefont {Arrachea}},
  \bibinfo {author} {\bibfnamefont {M.}~\bibnamefont {Moskalets}}, \ and\
  \bibinfo {author} {\bibfnamefont {D.}~\bibnamefont {S{\'{a}}nchez}},\ }\href
  {\doibase 10.3390/e18110419} {\bibfield  {journal} {\bibinfo  {journal}
  {Entropy}\ }\textbf {\bibinfo {volume} {18}},\ \bibinfo {pages} {419}
  (\bibinfo {year} {2016})}\BibitemShut {NoStop}%
\bibitem [{\citenamefont {Ludovico}\ \emph {et~al.}(2014)\citenamefont
  {Ludovico}, \citenamefont {Lim}, \citenamefont {Moskalets}, \citenamefont
  {Arrachea},\ and\ \citenamefont {S\'anchez}}]{LudovicoSanchez:14}%
  \BibitemOpen
  \bibfield  {author} {\bibinfo {author} {\bibfnamefont {M.~F.}\ \bibnamefont
  {Ludovico}}, \bibinfo {author} {\bibfnamefont {J.~S.}\ \bibnamefont {Lim}},
  \bibinfo {author} {\bibfnamefont {M.}~\bibnamefont {Moskalets}}, \bibinfo
  {author} {\bibfnamefont {L.}~\bibnamefont {Arrachea}}, \ and\ \bibinfo
  {author} {\bibfnamefont {D.}~\bibnamefont {S\'anchez}},\ }\href {\doibase
  10.1103/PhysRevB.89.161306} {\bibfield  {journal} {\bibinfo  {journal} {Phys.
  Rev. B}\ }\textbf {\bibinfo {volume} {89}},\ \bibinfo {pages} {161306}
  (\bibinfo {year} {2014})}\BibitemShut {NoStop}%
\bibitem [{\citenamefont {Eich}\ \emph {et~al.}(2016)\citenamefont {Eich},
  \citenamefont {Di~Ventra},\ and\ \citenamefont {Vignale}}]{EichVignale:16}%
  \BibitemOpen
  \bibfield  {author} {\bibinfo {author} {\bibfnamefont {F.~G.}\ \bibnamefont
  {Eich}}, \bibinfo {author} {\bibfnamefont {M.}~\bibnamefont {Di~Ventra}}, \
  and\ \bibinfo {author} {\bibfnamefont {G.}~\bibnamefont {Vignale}},\ }\href
  {\doibase 10.1103/PhysRevB.93.134309} {\bibfield  {journal} {\bibinfo
  {journal} {Phys. Rev. B}\ }\textbf {\bibinfo {volume} {93}},\ \bibinfo
  {pages} {134309} (\bibinfo {year} {2016})}\BibitemShut {NoStop}%
\bibitem [{\citenamefont {Tuovinen}\ \emph {et~al.}(2014)\citenamefont
  {Tuovinen}, \citenamefont {Perfetto}, \citenamefont {Stefanucci},\ and\
  \citenamefont {van Leeuwen}}]{TuovinenVanLeeuwen:14}%
  \BibitemOpen
  \bibfield  {author} {\bibinfo {author} {\bibfnamefont {R.}~\bibnamefont
  {Tuovinen}}, \bibinfo {author} {\bibfnamefont {E.}~\bibnamefont {Perfetto}},
  \bibinfo {author} {\bibfnamefont {G.}~\bibnamefont {Stefanucci}}, \ and\
  \bibinfo {author} {\bibfnamefont {R.}~\bibnamefont {van Leeuwen}},\ }\href
  {\doibase 10.1103/PhysRevB.89.085131} {\bibfield  {journal} {\bibinfo
  {journal} {Phys. Rev. B}\ }\textbf {\bibinfo {volume} {89}},\ \bibinfo
  {pages} {085131} (\bibinfo {year} {2014})}\BibitemShut {NoStop}%
\bibitem [{\citenamefont {Ridley}\ \emph {et~al.}(2015)\citenamefont {Ridley},
  \citenamefont {MacKinnon},\ and\ \citenamefont
  {Kantorovich}}]{RidleyKantorovich:15}%
  \BibitemOpen
  \bibfield  {author} {\bibinfo {author} {\bibfnamefont {M.}~\bibnamefont
  {Ridley}}, \bibinfo {author} {\bibfnamefont {A.}~\bibnamefont {MacKinnon}}, \
  and\ \bibinfo {author} {\bibfnamefont {L.}~\bibnamefont {Kantorovich}},\
  }\href {\doibase 10.1103/PhysRevB.91.125433} {\bibfield  {journal} {\bibinfo
  {journal} {Phys. Rev. B}\ }\textbf {\bibinfo {volume} {91}},\ \bibinfo
  {pages} {125433} (\bibinfo {year} {2015})}\BibitemShut {NoStop}%
\bibitem [{\citenamefont {Keldysh}(1964)}]{Keldysh:64_original}%
  \BibitemOpen
  \bibfield  {author} {\bibinfo {author} {\bibfnamefont {L.~V.}\ \bibnamefont
  {Keldysh}},\ }\href@noop {} {\bibfield  {journal} {\bibinfo  {journal} {Zh.
  Eksp. Teor. Fiz.}\ }\textbf {\bibinfo {volume} {47}},\ \bibinfo {pages}
  {1515} (\bibinfo {year} {1964})}\BibitemShut {NoStop}%
\bibitem [{\citenamefont {Keldysh}(1965)}]{Keldysh:65}%
  \BibitemOpen
  \bibfield  {author} {\bibinfo {author} {\bibfnamefont {L.~V.}\ \bibnamefont
  {Keldysh}},\ }\href@noop {} {\bibfield  {journal} {\bibinfo  {journal} {Sov.
  Phys. JETP}\ }\textbf {\bibinfo {volume} {20}},\ \bibinfo {pages} {1018}
  (\bibinfo {year} {1965})}\BibitemShut {NoStop}%
\bibitem [{\citenamefont {Stefanucci}\ and\ \citenamefont {van
  Leeuwen}(2013)}]{StefanucciVanLeeuwen:13}%
  \BibitemOpen
  \bibfield  {author} {\bibinfo {author} {\bibfnamefont {G.}~\bibnamefont
  {Stefanucci}}\ and\ \bibinfo {author} {\bibfnamefont {R.}~\bibnamefont {van
  Leeuwen}},\ }\href@noop {} {\emph {\bibinfo {title} {Nonequilibrium Many-Body
  Theory of Quantum Systems: A Modern Introduction}}}\ (\bibinfo  {publisher}
  {Cambridge University Press},\ \bibinfo {address} {Cambridge},\ \bibinfo
  {year} {2013})\BibitemShut {NoStop}%
\bibitem [{\citenamefont {Meir}\ and\ \citenamefont
  {Wingreen}(1992)}]{MeirWingreen:92}%
  \BibitemOpen
  \bibfield  {author} {\bibinfo {author} {\bibfnamefont {Y.}~\bibnamefont
  {Meir}}\ and\ \bibinfo {author} {\bibfnamefont {N.~S.}\ \bibnamefont
  {Wingreen}},\ }\href {\doibase 10.1103/PhysRevLett.68.2512} {\bibfield
  {journal} {\bibinfo  {journal} {Phys. Rev. Lett.}\ }\textbf {\bibinfo
  {volume} {68}},\ \bibinfo {pages} {2512} (\bibinfo {year}
  {1992})}\BibitemShut {NoStop}%
\bibitem [{\citenamefont {Runge}\ and\ \citenamefont
  {Gross}(1984)}]{RungeGross:84}%
  \BibitemOpen
  \bibfield  {author} {\bibinfo {author} {\bibfnamefont {E.}~\bibnamefont
  {Runge}}\ and\ \bibinfo {author} {\bibfnamefont {E.~K.~U.}\ \bibnamefont
  {Gross}},\ }\href@noop {} {\bibfield  {journal} {\bibinfo  {journal} {Phys.
  Rev. Lett.}\ }\textbf {\bibinfo {volume} {52}},\ \bibinfo {pages} {997}
  (\bibinfo {year} {1984})}\BibitemShut {NoStop}%
\bibitem [{\citenamefont {Ullrich}(2012)}]{Ullrich:12}%
  \BibitemOpen
  \bibfield  {author} {\bibinfo {author} {\bibfnamefont {C.~A.}\ \bibnamefont
  {Ullrich}},\ }\href@noop {} {\emph {\bibinfo {title} {{T}ime-{D}ependent
  {D}ensity-{F}unctional {T}heory: {C}oncepts and {A}pplications}}},\ Oxford
  Graduate Texts\ (\bibinfo  {publisher} {Oxford University Press},\ \bibinfo
  {address} {Oxford},\ \bibinfo {year} {2012})\BibitemShut {NoStop}%
\bibitem [{\citenamefont {Kurth}\ \emph {et~al.}(2005)\citenamefont {Kurth},
  \citenamefont {Stefanucci}, \citenamefont {Almbladh}, \citenamefont {Rubio},\
  and\ \citenamefont {Gross}}]{KurthGross:05}%
  \BibitemOpen
  \bibfield  {author} {\bibinfo {author} {\bibfnamefont {S.}~\bibnamefont
  {Kurth}}, \bibinfo {author} {\bibfnamefont {G.}~\bibnamefont {Stefanucci}},
  \bibinfo {author} {\bibfnamefont {C.-O.}\ \bibnamefont {Almbladh}}, \bibinfo
  {author} {\bibfnamefont {A.}~\bibnamefont {Rubio}}, \ and\ \bibinfo {author}
  {\bibfnamefont {E.~K.~U.}\ \bibnamefont {Gross}},\ }\href {\doibase
  10.1103/PhysRevB.72.035308} {\bibfield  {journal} {\bibinfo  {journal} {Phys.
  Rev. B}\ }\textbf {\bibinfo {volume} {72}},\ \bibinfo {pages} {035308}
  (\bibinfo {year} {2005})}\BibitemShut {NoStop}%
\bibitem [{\citenamefont {Stefanucci}\ \emph {et~al.}(2007)\citenamefont
  {Stefanucci}, \citenamefont {Kurth}, \citenamefont {Gross},\ and\
  \citenamefont {Rubio}}]{StefanucciRubio:07}%
  \BibitemOpen
  \bibfield  {author} {\bibinfo {author} {\bibfnamefont {G.}~\bibnamefont
  {Stefanucci}}, \bibinfo {author} {\bibfnamefont {S.}~\bibnamefont {Kurth}},
  \bibinfo {author} {\bibfnamefont {E.}~\bibnamefont {Gross}}, \ and\ \bibinfo
  {author} {\bibfnamefont {A.}~\bibnamefont {Rubio}},\ }in\ \href {\doibase
  https://doi.org/10.1016/S1380-7323(07)80028-8} {\emph {\bibinfo {booktitle}
  {Molecular and Nano Electronics:Analysis, Design and Simulation}}},\ \bibinfo
  {series} {Theoretical and Computational Chemistry}, Vol.~\bibinfo {volume}
  {17},\ \bibinfo {editor} {edited by\ \bibinfo {editor} {\bibfnamefont
  {J.}~\bibnamefont {Seminario}}}\ (\bibinfo  {publisher} {Elsevier},\ \bibinfo
  {year} {2007})\ pp.\ \bibinfo {pages} {247 -- 284}\BibitemShut {NoStop}%
\bibitem [{\citenamefont {Kurth}\ and\ \citenamefont
  {Stefanucci}(2013)}]{KurthStefanucci:13}%
  \BibitemOpen
  \bibfield  {author} {\bibinfo {author} {\bibfnamefont {S.}~\bibnamefont
  {Kurth}}\ and\ \bibinfo {author} {\bibfnamefont {G.}~\bibnamefont
  {Stefanucci}},\ }\href {\doibase 10.1103/PhysRevLett.111.030601} {\bibfield
  {journal} {\bibinfo  {journal} {Phys. Rev. Lett.}\ }\textbf {\bibinfo
  {volume} {111}},\ \bibinfo {pages} {030601} (\bibinfo {year}
  {2013})}\BibitemShut {NoStop}%
\bibitem [{\citenamefont {Eich}\ \emph
  {et~al.}(2014{\natexlab{b}})\citenamefont {Eich}, \citenamefont {Di~Ventra},\
  and\ \citenamefont {Vignale}}]{EichVignale:14a}%
  \BibitemOpen
  \bibfield  {author} {\bibinfo {author} {\bibfnamefont {F.~G.}\ \bibnamefont
  {Eich}}, \bibinfo {author} {\bibfnamefont {M.}~\bibnamefont {Di~Ventra}}, \
  and\ \bibinfo {author} {\bibfnamefont {G.}~\bibnamefont {Vignale}},\ }\href
  {\doibase 10.1103/PhysRevLett.112.196401} {\bibfield  {journal} {\bibinfo
  {journal} {Phys. Rev. Lett.}\ }\textbf {\bibinfo {volume} {112}},\ \bibinfo
  {pages} {196401} (\bibinfo {year} {2014}{\natexlab{b}})}\BibitemShut
  {NoStop}%
\bibitem [{\citenamefont {Eich}\ \emph {et~al.}(2017)\citenamefont {Eich},
  \citenamefont {Ventra},\ and\ \citenamefont {Vignale}}]{EichVignale:17a}%
  \BibitemOpen
  \bibfield  {author} {\bibinfo {author} {\bibfnamefont {F.~G.}\ \bibnamefont
  {Eich}}, \bibinfo {author} {\bibfnamefont {M.~D.}\ \bibnamefont {Ventra}}, \
  and\ \bibinfo {author} {\bibfnamefont {G.}~\bibnamefont {Vignale}},\ }\href
  {http://stacks.iop.org/0953-8984/29/i=6/a=063001} {\bibfield  {journal}
  {\bibinfo  {journal} {Journal of Physics: Condensed Matter}\ }\textbf
  {\bibinfo {volume} {29}},\ \bibinfo {pages} {063001} (\bibinfo {year}
  {2017})}\BibitemShut {NoStop}%
\end{thebibliography}%

\end{document}